\documentclass[superscriptaddress,preprint,12pt]{revtex4-1}
\usepackage{xcolor}
\usepackage{rotating}
\usepackage{amsmath}
\usepackage{bm}
\usepackage{array}
\usepackage{graphicx}
\usepackage{dcolumn}   
\usepackage{amssymb}
\usepackage{mathtools}
\usepackage[english]{babel}

\begin{document}

\title{Method of regularised stokeslets: \\Flow analysis and improvement of convergence}
\author{Boan Zhao}
\affiliation{ Department of Applied Mathematics and Theoretical Physics, University of Cambridge, Wilberforce Road, Cambridge CB3 0WA, United Kingdom}
\author{Eric Lauga}
\affiliation{ Department of Applied Mathematics and Theoretical Physics, University of Cambridge, Wilberforce Road, Cambridge CB3 0WA, United Kingdom}
\author{Lyndon Koens\footnote{lyndon.koens@mq.edu.au}}
\affiliation{ Department of Applied Mathematics and Theoretical Physics, University of Cambridge, Wilberforce Road, Cambridge CB3 0WA, United Kingdom}
\affiliation{ Department of Mathematics and Statistics, Macquarie University, 192 Balaclava Rd, Macquarie Park, NSW 2113, Australia}

\begin{abstract}
Since their development in 2001, regularised stokeslets have become a popular numerical tool for low-Reynolds number flows  since the replacement of a point force by a smoothed blob overcomes many computational difficulties associated with flow singularities (Cortez, 2001, \textit{SIAM J. Sci. Comput.} \textbf{23}, 1204). The physical changes to the flow resulting from this process are, however, unclear. In this paper, we analyse the flow induced by general regularised stokeslets. An explicit formula for the flow from any regularised stokeslet is first derived, which is shown to simplify for spherically symmetric blobs. Far from the centre of any regularised stokeslet we show that the flow can be written in terms of an infinite number of singularity solutions provided  the blob decays sufficiently rapidly. This infinite number of singularities reduces to a point force and source dipole for spherically symmetric blobs. Slowly-decaying blobs induce additional flow resulting from the non-zero body forces acting on the fluid. We also show that near the centre of spherically symmetric regularised stokeslets the flow becomes isotropic, which  contrasts with the flow anisotropy   fundamental to viscous systems. The concepts developed are used to { identify blobs that reduce regularisation errors. These blobs contain regions of negative force in order to counter the flows produced in the regularisation process, but still retain a form convenient for computations. }
\end{abstract}
\maketitle

\section{Introduction}

Viscous flows play key roles in microscopic biology \cite{Nazockdast2017, Lauga2009}, colloidal science \cite{DuRoure2019,Cates2018}, and engineering \cite{Inglis2006,Zhang2019}. In these flows the ratio of fluid inertia to viscous dissipation, the Reynolds number, is very small, and thus the dynamics is essentially inertia-less. Yet these flows display many surprising and emergent phenomena. For example, microscopic organisms, such as bacteria \cite{Lauga2016} and spermatozoa \cite{Gaffney2011}, have developed methods to robustly swim in this viscous world. This swimming is distinct from our own locomotion which relies on inertia. The principles of viscous swimming can be critical to several body functions as collections of microscopic beating filaments, called cilia, exist throughout our bodies to clear passages and airways \cite{Golestanian2011}. Beyond the biological world, even simple suspensions can be complex \cite{Sasayama2017, DuRoure2019, Tornberg2004}. For example, the motion of a single rigid fibre in viscous flows is fairly well understood \cite{Jeffery1922} but suspensions of fibres create surprising and significant changes to the bulk rheology \cite{DuRoure2019} and conductivity \cite{Beckers2015} of the fluid.

A dedicated combination of experiments and theory is often needed to elucidate these phenomena. However each of these explorations pose their own challenges. Many of the situations of interest are microscopic in size \cite{KimKarrila}, making direct experimental manipulation and observation difficult. The development of novel micro-machines may help with this in the coming years \cite{Zhang2019, Koens2019, Qiu2015}. These issues become even more complex in biological systems, as changes in temperature, chemical compositions, pH, and light can induce undesired responses \cite{Goldstein2015,Young2006,Nanninga1998}. Genetic modification could remove these behaviours \cite{Aswad1975,Shah2000,Lockman1990,Goldstein2015} but often has unintended effects. 

Theoretical explorations of viscous flow are, in contrast, more established. Viscous flows are well described using the incompressible Stokes equations \cite{KimKarrila}. The dynamics of a system can, in principle, be found by solving these equations with the appropriate boundary conditions. However, exact solutions for these equations only exist for simple shapes like spheres or ellipsoids \cite{ChwangWu} and so in general solutions must be sought numerically \cite{Pozrikidis1992}. Even so, these equations have been successfully used to model, for example and among many others,  the sedimentation of rods \cite{Tornberg2004}, suspensions of spheres \cite{Ekiel-Jeewska2009}, the flows inside biological cells \cite{Nazockdast2017,Goldstein2015a}, the drag on dandelion seeds  \cite{Cummins2018}, the diffusion of bacteria \cite{Koens2014,Ramia1993} and the dynamics of micro-machines \cite{Koens2019,Montenegro-Johnson2016}. Fundamental rules of viscous swimming have also been revealed \cite{Purcell,BECKER2003,Koens2016a} and asymptotic tools have been developed to improve their computation \cite{GRAY1955,Batchelor2006,1976,Keller1976a,Johnson1979,Koens2016,Koens2018}.

In the 20th century, many theoretical explorations of Stokes flow relied on the flow from a point force, called the stokeslet \cite{KimKarrila}. This   is the Green's function for Stokes flow, and it allows higher-order singularity solutions, such as  force dipoles, to be directly constructed. As a result, any solution to Stokes flow could be represented by a combination of singularity solutions, often called the representation by fundamental singularities \cite{ChwangWu}, or an integral of point forces over the surface of the body, called the boundary integral representation \cite{Pozrikidis1992}. Unfortunately, the flow generated from a point force diverges at the location of the force  and thus representations by fundamental singularities require point singularities to not be placed in the flow region (i.e.~they must be virtual singularities and placed within the walls) \cite{ChwangWu} while boundary integrals have to deal with the integral of a divergent function \cite{Pozrikidis1992} and thus numerical calculations require additional care to avoid any issues.

\begin{figure}
\centering
\includegraphics[height=130pt]{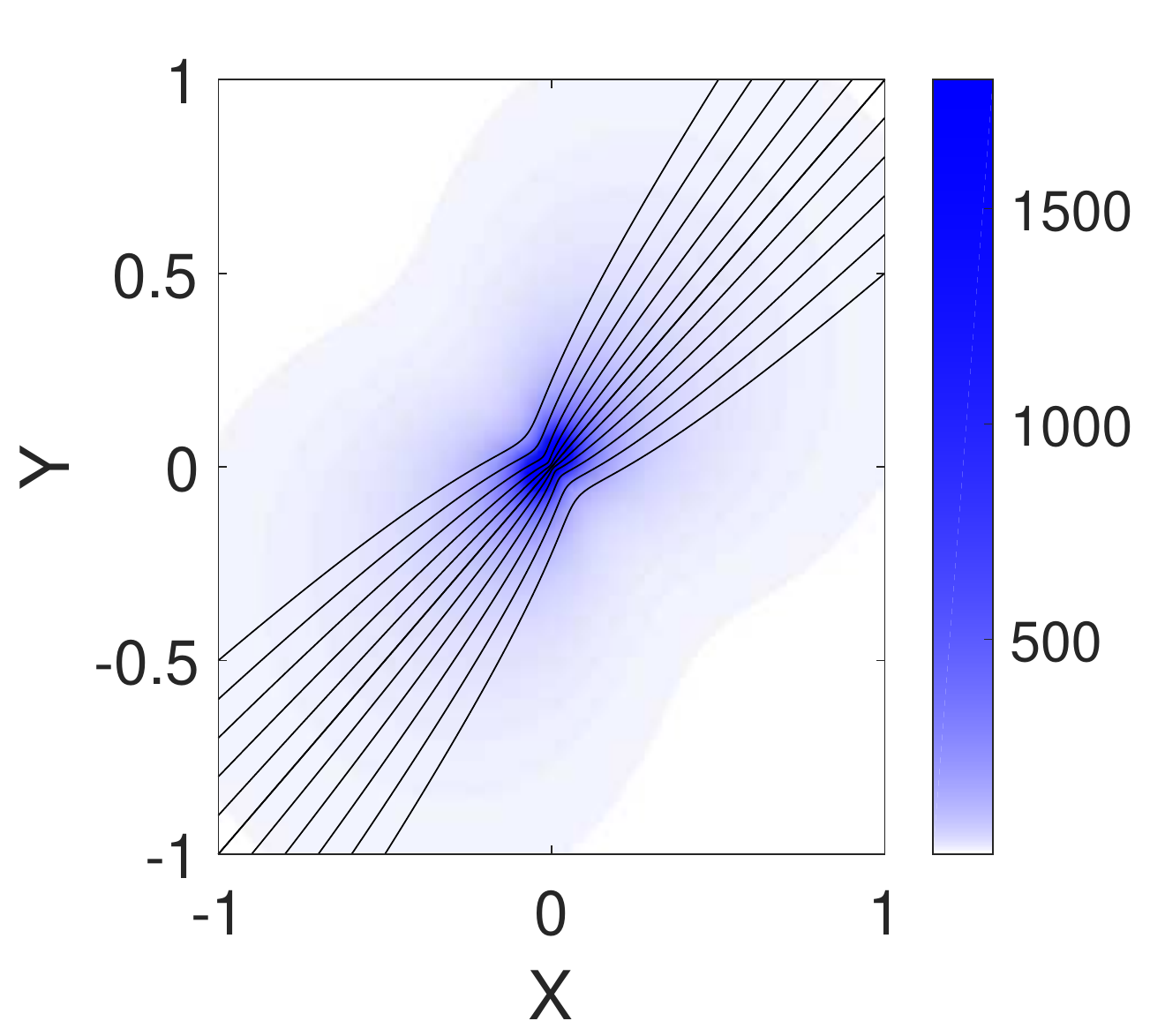}
\includegraphics[height=130pt]{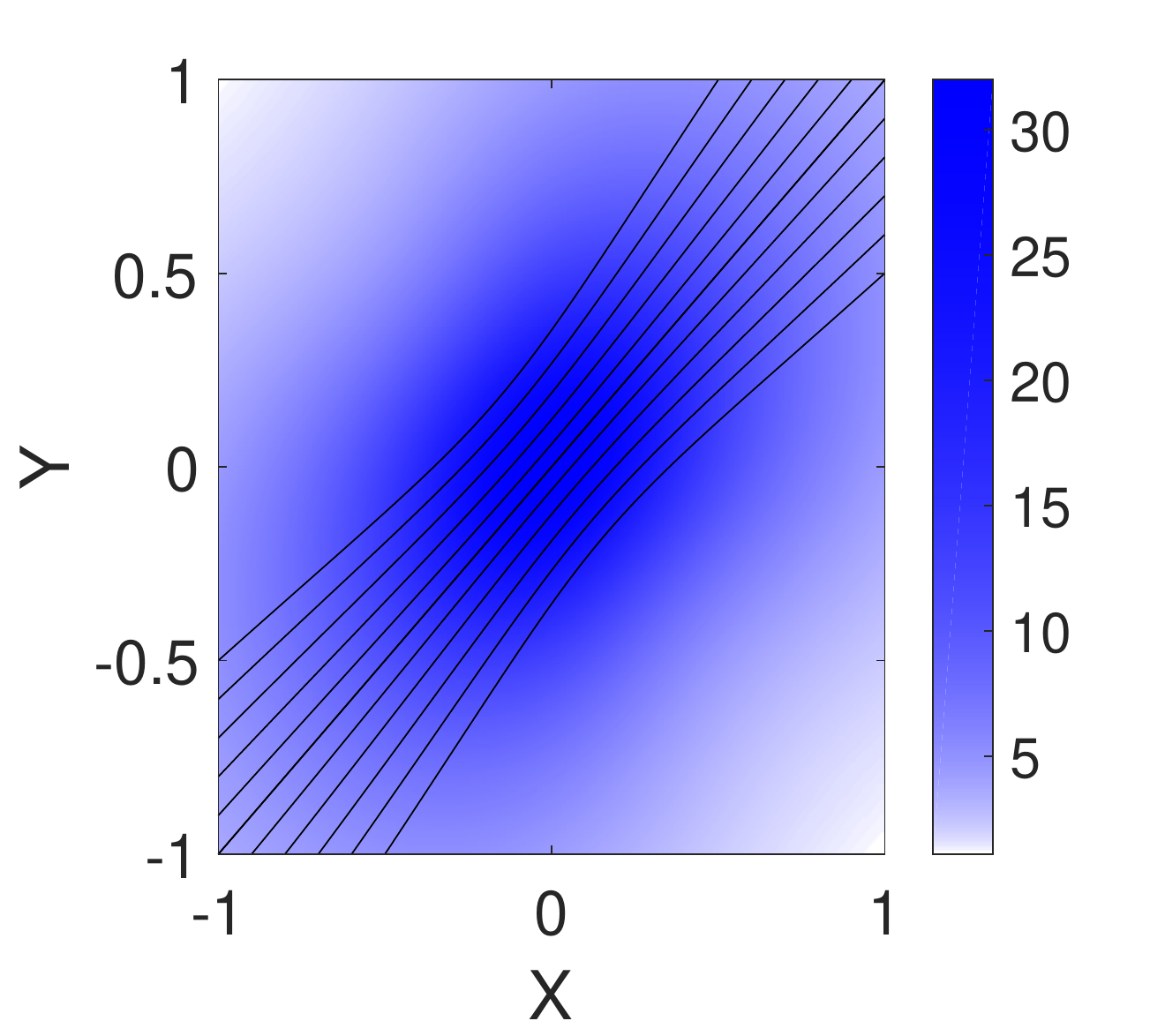}
\includegraphics[height=130pt]{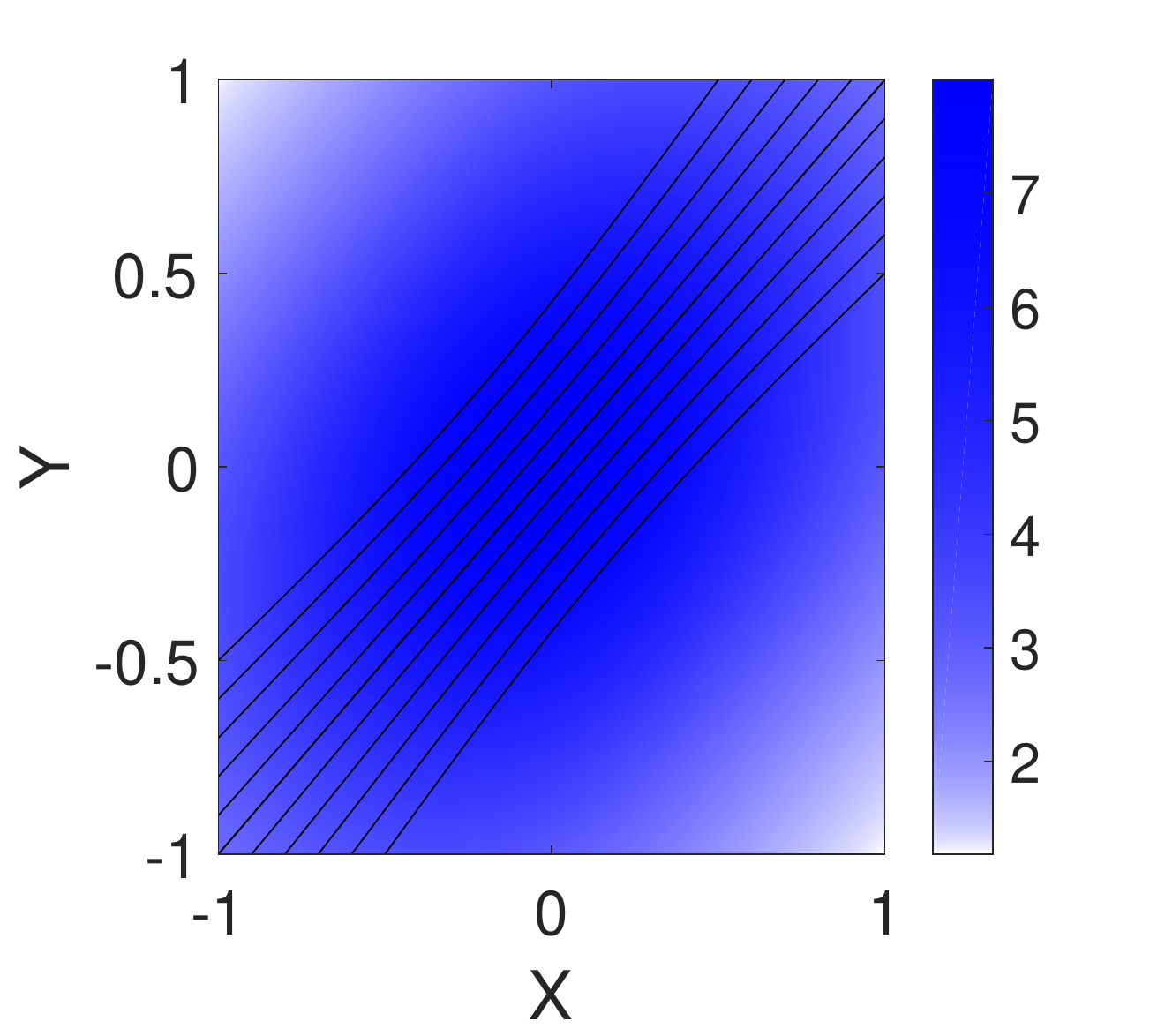}
\caption{Streamlines (lines) and flow strength (density plot) from an example of a regularised stokeslet with $\epsilon=0$ (left), $\epsilon=0.5$ (middle) and $\epsilon=1$ (right). The net dimensionless force in each case is $\mathbf{F}=(1,1,0)$. The regularised stokeslet has the algebraic blob $f(r) =15/(8\pi\sqrt{1+r^2})$. { The units for $X$ and $Y$ are arbitrary.}.} 
\label{figure_Comparison}
\end{figure}

In 2001, Cortez devised the method of regularised stokeslets to overcome these computational issues \cite{Cortez2001}. This method replaced the point forces in the traditional solution methods with finite-size blobs of spatially-distributed force. The flow from one such a blob of force is finite everywhere and so effectively removes the divergent flow in the point force. Cortez termed the flow solutions resulting from these blobs  regularised stokeslets. The size of the blobs used were characterised by a parameter $\epsilon$ and chosen such that the flow would limit mathematically to the stokeslet flow as $\epsilon\rightarrow 0$ (Fig.~\ref{figure_Comparison}). This ensures that the solution found is similar to the actual flow but with an additional error related to $\epsilon$ \cite{Cortez2005}. Small regularisation parameters, $\epsilon$, create  flows close to the real solution and are free of any divergence issues. This makes numerical simulations of Stokes flow using regularised stokeslets easy to implement \cite{Smith2018} and so they have become very popular. For example, regularised stokeslets have recently been  used to study microscopic swimming \cite{Godinez2015, Ishimoto2018}, phoretic flows \cite{Montenegro-Johnson2015, Montenegro-Johnson2018}, flexible structures \cite{Olson2013} and the pumping from cilla \cite{Montenegro-Johnson2012}. These results typically compare favourably with exact solutions and experiments. The method has also been extended to consider slender-filaments \cite{Smith, HoaNguyen2014, Montenegro-Johnson2016, Cortez2012, Rodenborn2013a, Bouzarth}, flows near walls \cite{Ainley2008,Cortez2015} and Brinkman fluids \cite{Nguyen2018}. 

The accuracy of the regularised stokeslet method is directly related to the flow generated from each blob. Regularisation blobs that quickly recover the stokeslet flow as $\epsilon \rightarrow 0$ are inherently more accurate mathematically. Yet the types of flows produced by a regularisation blob and the rate at which the flow returns to the stokeslet are generally not considered. In this  paper, we analyse the flow from general regularised stokeslets in order to provide relationships between the regularisation and nature of the flow field. This, in turn, reveals the accuracy of different regularisations. We provide an explicit equation for the flow from a general regularisation blob with no assumed symmetries and analyse how the flow behaves. Far from the centre of the regularisation, the flow from exponentially decaying blobs can be represented by a combination of singularity solutions (point force, force dipole, etc.) while blobs with power-law decays induce additional flow from the non-zero body force on the fluid. Apart from the stokeslet contribution, each of these flows are proportional to a power of $\epsilon$. The number of terms depends on the symmetries of the blob and reduces to two singularities for spherically-symmetric rapidly-decaying blobs. { The far-field contribution from this second singularity cannot be removed if the regularisation is always greater than 0 (strictly positive).} The full flow from a spherically symmetric regularised stokeslets is also found to have a simple representation in terms of integrals of the regularisation blob and the two singularity solutions. This reveals in particular that, near the centre of the blobs, the flow becomes isotropic, in contrast with the anisotropy inherent in the stokeslet. The contribution of this isotropic region to surface and line integrals is estimated. Finally all these results are used to develop two regularised stokeslets that converge exponentially, at least,  to the flow from a stokeslet. These minimise the error produced by regularisation while keeping a form which is computationally simple.

The paper is organised as follows. Section~\ref{sec:reg_stokes} provides a mathematical background into singularity solutions of Stokes flow and regularised singularities. In particular it describes the general regularisation blobs considered and introduces three example blobs: a power-law blob, a Gaussian blob and a compactly supported blob. The power-law and Gaussian blobs are popular examples, while the compactly supported blob is new to the our knowledge. In Sec.~\ref{sec:taylor} we expand the examples in a Taylor series around $\epsilon=0$ to determine their far-field flows before an expansion of a general regularised stokeslet is produced in Sec.~\ref{sec:convolution}. A compact representation for the flow from a spherically symmetric stokeslets is then developed in Sec.~\ref{sec:sphere} and is used to explore the flow near and far from the centre of these blobs. Sec.~\ref{sec:good} uses these results to produce two exponentially accurate regularised stokeslets. The results are summarised in Sec.~\ref{sec:conclusion}.

\section{Singularities and regularisation} \label{sec:reg_stokes} 

Viscous flows at low-Reynolds number  are well described by the incompressible Stokes equations \cite{KimKarrila}
\begin{eqnarray}
-\nabla p+\mu \nabla^2\mathbf{u} +\mathbf{f} &=& \mathbf{0} ,\\
\nabla\cdot\mathbf{u} &=& 0,
\end{eqnarray}
where $p$ is the dynamic pressure, $\mathbf{u}$ is the fluid velocity, $\mu$ is the viscosity of the fluid, and $\mathbf{f}$ is the force per unit volume on the fluid. The viscosity is set to $\mu=1$ for the rest of this paper. Since these equations are time independent, the flow has no memory and so only depends on the geometry of a system at a given instant. The equations are also linear, which  means any flow can be represented through a combination of simpler flow solutions, such as uniform flows, shear flow, the flow from a point force (stokeslet) and their derivatives \cite{KimKarrila}. The stokeslet and its derivatives are called singularity solutions to the Stokes equations and each have a physical interpretation. The stokeslet is the flow from a point force, the derivative of a stokeslet is the flow from a force dipole, etc. The interpretation of regularised singularities are not as clear because of the replacement of a point with a blob. This section provides a background on the singularity solutions to Stokes flow and the derivation of the regularised stokeslet.

\subsection{Background: Singularity solutions to Stokes flow}

\begin{figure}
\centering
\includegraphics[height=200pt,width = 200pt]{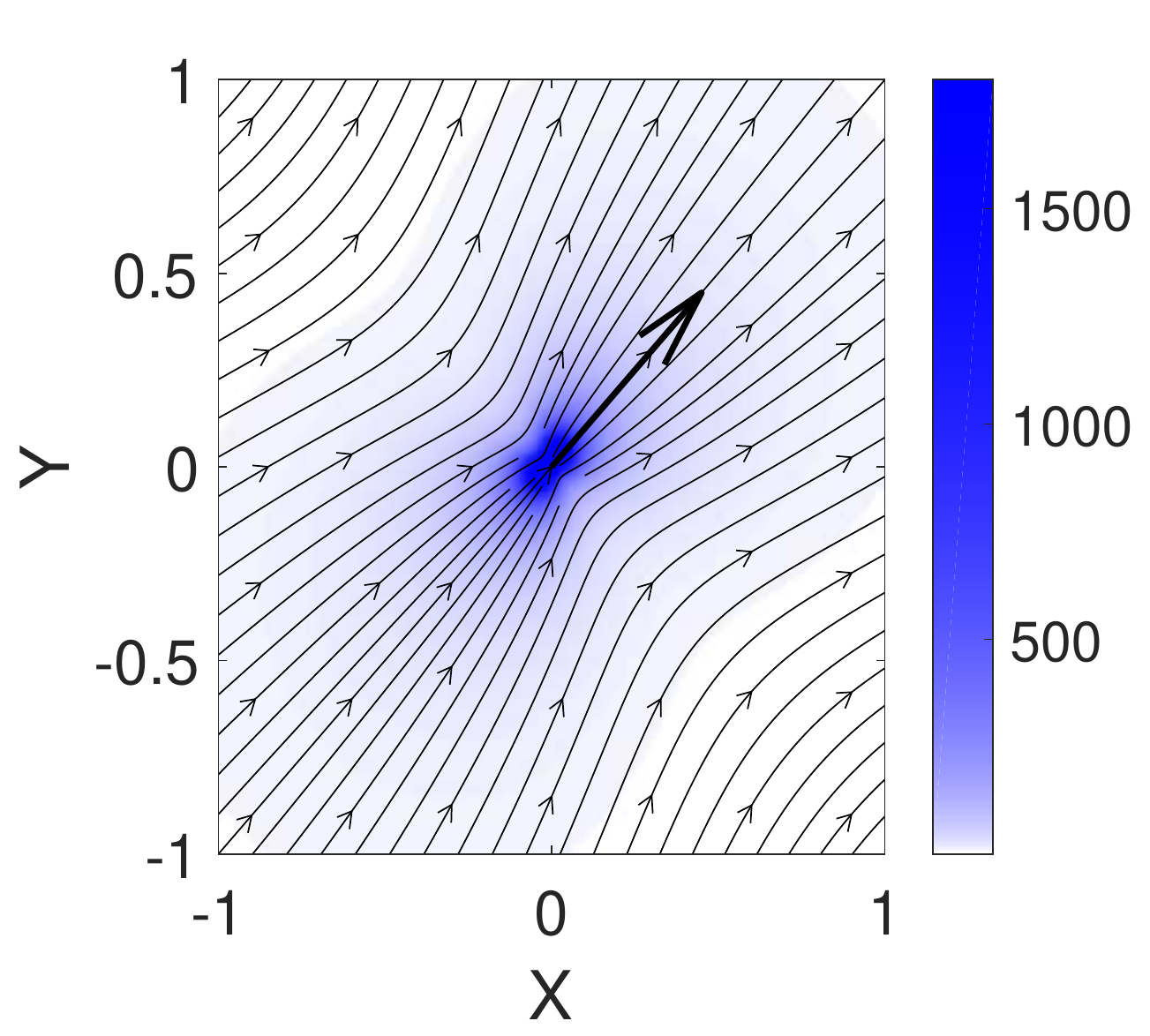}
\includegraphics[height=200pt,width = 200pt]{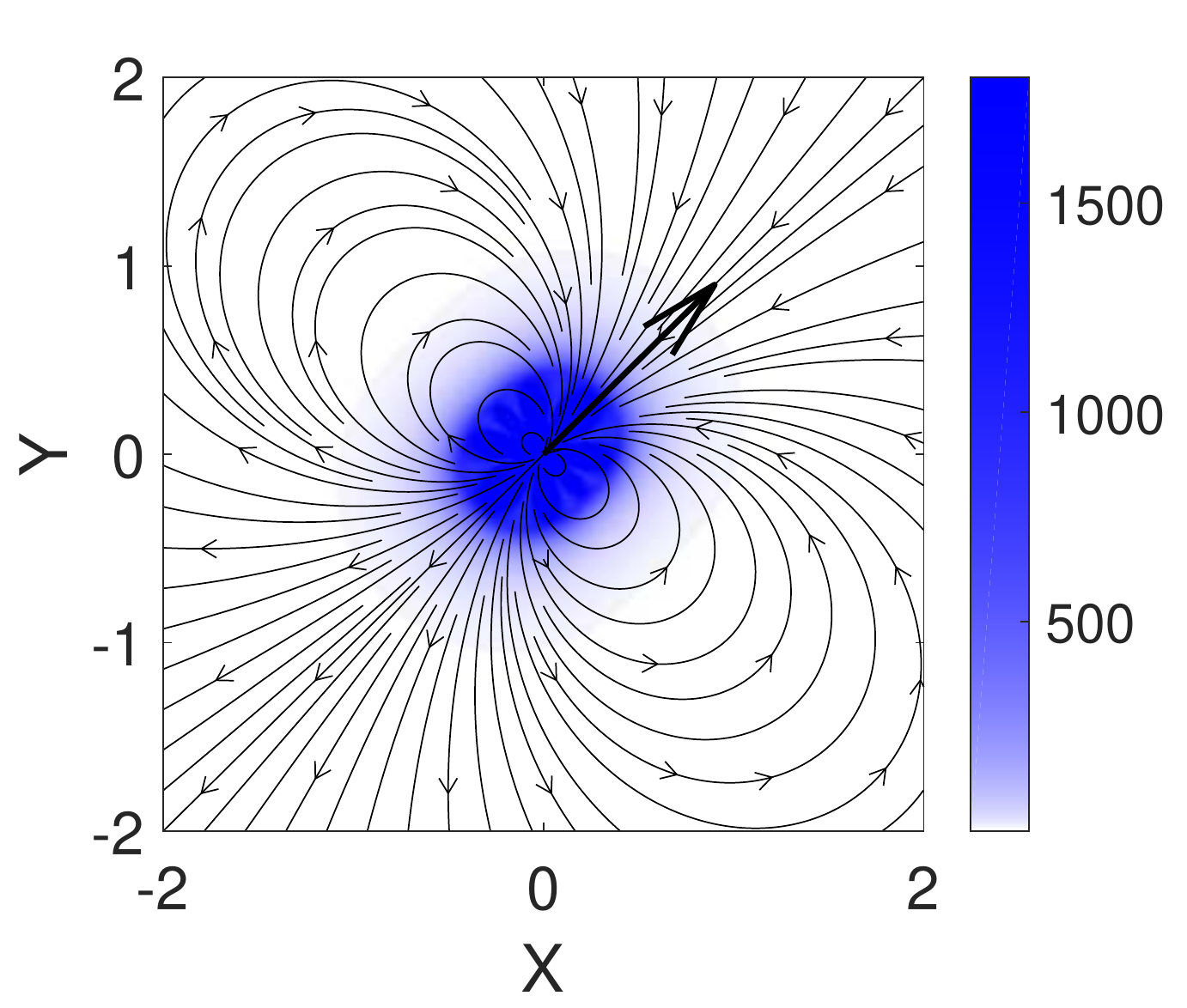}
\caption{Streamlines (black lines) and velocity speed (density plot) of a stokeslet (left) and source dipole (right) solutions aligned with the arrows. The stokeslet has dimensionless strength $\mathbf{F}=(1,1,0)$ and the source dipole has the sink placed to the top right of the source. { The units for $X$ and $Y$ are arbitrary.} }
\label{figure_Stokeslet}
\end{figure}

The fundamental flow singularity is called the stokeslet, and represents the flow from a point force in the fluid (Fig.~\ref{figure_Stokeslet}, left). This singularity is found when $\mathbf{f}=\mathbf{F}\delta(\mathbf{r})$ and produces the flow 
\begin{eqnarray}
\mathbf{u}_F(\mathbf{r}) = \mathbf{S}(\mathbf{r})\cdot\mathbf{F} &=&
\left(\frac{r^2\mathbf{I}+\mathbf{rr}}{8\pi 
r^3}\right)\cdot\mathbf{F},\\
 p_F(\mathbf{r}) =
\mathbf{P}(\mathbf{r})\cdot\mathbf{F} &=&
\left(\frac{\mathbf{r}}{4\pi r^3}\right)\cdot\mathbf{F},
\end{eqnarray}
where $\mathbf{F}$ is the vector of the point force, $\delta(\mathbf{r})$ is the Dirac delta function centred at the origin, $\mathbf{u}_F(\mathbf{r}) $ is the velcoity of the fluid from the stokeslet  the location $\mathbf{r}$,  $p_F(\mathbf{r})$ is the pressure from the stokeslet at $\mathbf{r}$, $\mathbf{S}(\mathbf{r})$ is the Oseen tensor, $\mathbf{P}(\mathbf{r})$ is the corresponding  pressure tensor, $r=|\mathbf{r}|$ is the length of $\mathbf{r}$, $\mathbf{I}$ is the identity tensor, and all tensor product symbols are omitted. The above flow assumes the fluid is at rest far from the point force and diverges when $r = 0$. 

The flow from a stokeslet can be written as the Oseen tensor contracted with $\mathbf{F}$.  This contracted tensor structure is common to all singularity flows, with the isolated tensor called the flow tensor of the singularity. These flow tensors contain the behaviour of the flow for all possible configurations along the different `columns' of the tensor. Hence the flow tensors will be effective for analysing the regularised stokeslet. The Oseen and the corresponding pressure tensor satisfy
\begin{eqnarray} \label{eqn_Stokeslet_Definition}
-\nabla \mathbf{P}+ \nabla^2\mathbf{S}+\delta(\mathbf{r})\mathbf{I}&=& \mathbf{0},\\
\nabla\cdot\mathbf{S} &=& 0,
\end{eqnarray} 
where the columns of $\mathbf{S}$ represent forcing along different basis vectors. 

Another equally important singularity for Stokes flow is the flow from a point source of fluid located at the origin. This irrotational flow  has the form
\begin{equation}
\mathbf{u}_{src}(\mathbf{r}) = \frac{\mathbf{r}}{4\pi r^3} q,
\end{equation}
where $q$ is the strength of the point source. If the strength of the source, $q$, is negative this flow represents a point sink. This singularity is important when the volume of the bodies can change, such as growing bubbles or droplets, and the higher-order singularities developed from it occur frequently alongside the point force \cite{ChwangWu}.

Higher-order singularity solutions can be constructed by differentiating and taking linear combinations of the two fundamental solutions above (stokeslet and point source). For example, the source dipole is formed by placing a source next to a sink of equal strength and taking the limit that the separation goes to zero while keeping the product of the separation and the strength constant. The flow tensor for the source dipole is  
\begin{equation}
\mathbf{D(r)} = \frac{r^2\mathbf{I}-3\mathbf{rr}}{8\pi
r^5} = \frac{1}{2q}\nabla\mathbf{u}_{src} =
\frac{1}{2}\nabla^2\mathbf{S}, \label{dipole}
\end{equation}
where $\mathbf{D(r)}$ is the source dipole tensor. The associated flow is depicted in Fig.~\ref{figure_Stokeslet}, right. The columns of $\mathbf{D(r)}$ correspond to different alignments of source and sink. The source dipole produces no flux through any sphere centred around the origin and frequently occurs in combination with the point force.

 The flow from a force dipole is found through the derivative of the stokeslet, similarly to the source dipole. Its flow tensor is a rank 3 tensor in order to account for both the direction of the force and alignments of the pair of forces. Normally the force dipole is decomposed into a symmetric component called the stresslet and an antisymmetric component called the rotlet \cite{KimKarrila}. The former of these represents pure strain flow while the latter is the flow from a point torque.

\subsection{Regularised singularities}

\begin{figure}
\centering
\includegraphics[height=180pt]{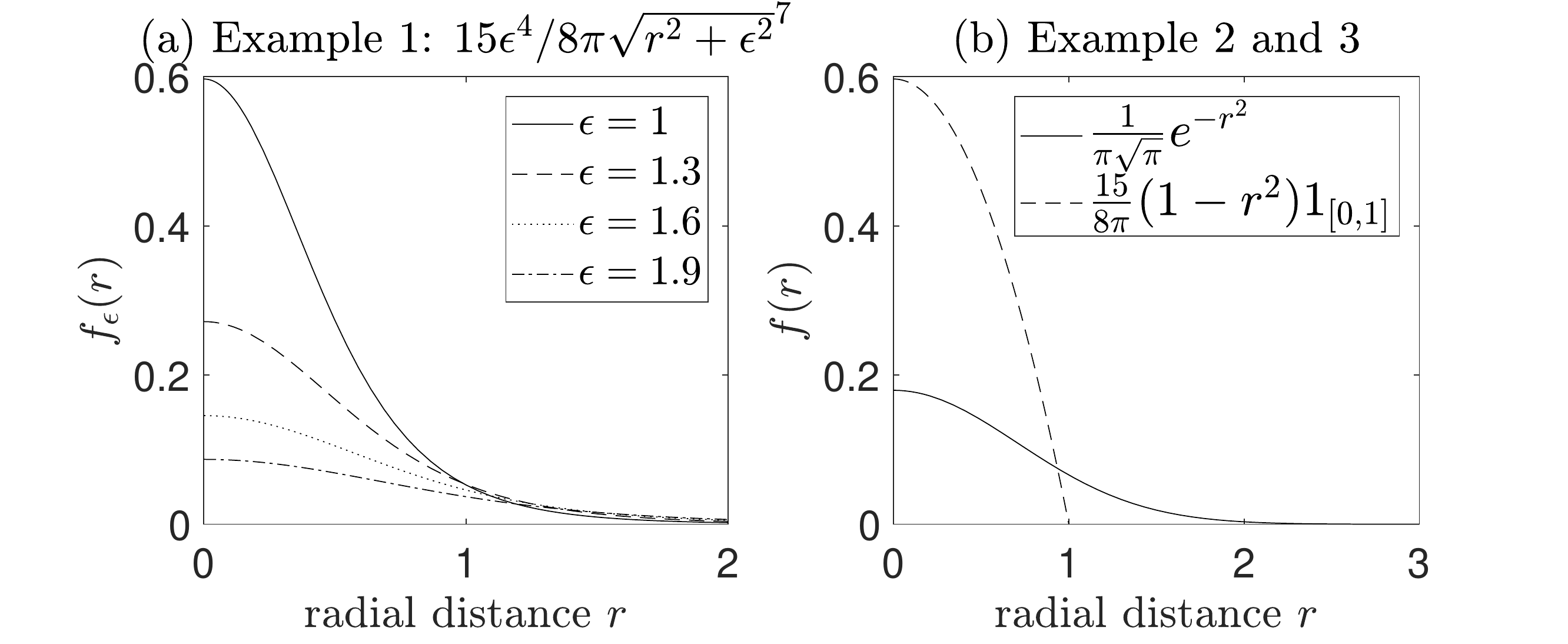}
\caption{Plots of the three blobs (a) The original blob used by Cortez,
plotted with different $\epsilon$ \cite{Cortez2001}. (b) Plot of the Gaussian and a blob of compact support. { All length have been scaled by $\epsilon$ as it is the only relevant length scale.}}
\label{figure_blobs}
\end{figure}

In his original article, Cortez regularised the singularity solutions to overcome their divergent flows when $|\mathbf{r}| = 0$ \cite{Cortez2001}. This provided computational simplicity and was achieved by replacing the Dirac delta in Eq.~\eqref{eqn_Stokeslet_Definition} with a family of mollifiers (hereafter called blobs) of the form
 \begin{equation}
f_\epsilon(\mathbf{r}) =
\epsilon^{-3}f(\mathbf{r}/\epsilon),
\end{equation}
where $f(\mathbf{r})$ is a general function, $\epsilon$ characterises the blob size and 
\begin{equation}
\lim_{\epsilon \to 0} f_\epsilon(\mathbf{r}) = \delta(\mathbf{r}).
\end{equation}
Blobs with the above properties are called approximations to the identity \cite{EliasRami} and by construction are normalised such that
\begin{eqnarray}
\iiint f(\mathbf{r})\,d\mathbf{r} = 1,
\end{eqnarray}
where the integral has been performed over all space. This normalisation condition is satisfied when $f(\mathbf{r})$ decays as $1/r^4$ or faster as $r \rightarrow \infty$. Many different blobs can satisfy these conditions. Some examples include the original power-law blob proposed by Cortez \cite{Cortez},
\begin{equation}
f_{\epsilon}^{p}(r) = \frac{15\epsilon^4}{8\pi r_\epsilon^7}, \label{blob_power}
\end{equation}
where $r_\epsilon=\sqrt{r^2+\epsilon^2}$, a Gaussian blob of
\begin{equation}
f_{\epsilon}^{g}(r)=\frac{1}{\pi\sqrt{\pi}\epsilon^3}e^{-r^2/\epsilon^2}, \label{blob_guassian}
\end{equation}
and a blob with compact support
\begin{equation}
f_{\epsilon}^{s}(r) = \left\{\begin{array}{c c}
\displaystyle \frac{15}{8\pi\epsilon^3}\left(1-\frac{r^2}{\epsilon^2}\right), & r<\epsilon \\
0 ,& r> \epsilon
\end{array} \right. . \label{blob_support}
\end{equation}
These blobs are illustrated in Fig.~\ref{figure_blobs}. The power-law blob, Eq.~\eqref{blob_power}, and the Gaussian blob, Eq.~\eqref{blob_guassian}, are both popular examples used within the literature \cite{Cortez2015}. The compactly supported blob, Eq.~\eqref{blob_support}, is less common and concentrates all the force in a ball of radius $\epsilon$. { The flow compact blobs are continuous but may have discontinuous derivatives. This means their derivatives do not always produce continuous higher order singularities. However, integrals of compact regularised stokeslets are well defined and so can be readily used in numerics.} Note that the function, $f(\mathbf{r})$, and the regularisation parameter, $\epsilon$, can be scaled simultaneously to generate the same family of functions. In what follows, we fix the function $f(\mathbf{r})$ and tune  the value of $\epsilon$   to avoid confusion (Fig.~\ref{figure_blobs}a). 

\begin{figure}
\centering
\includegraphics[height=190pt]{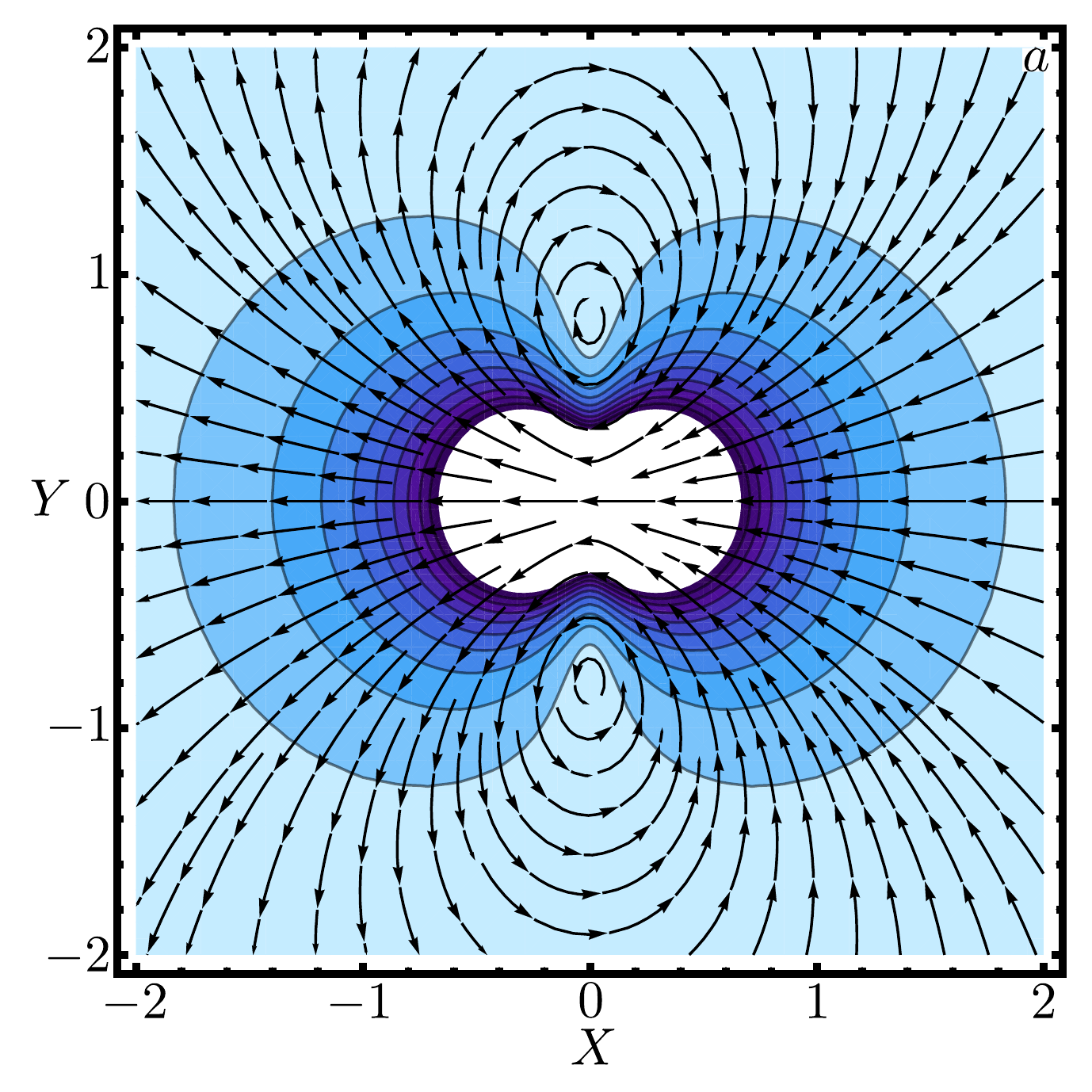}
\includegraphics[height=190pt]{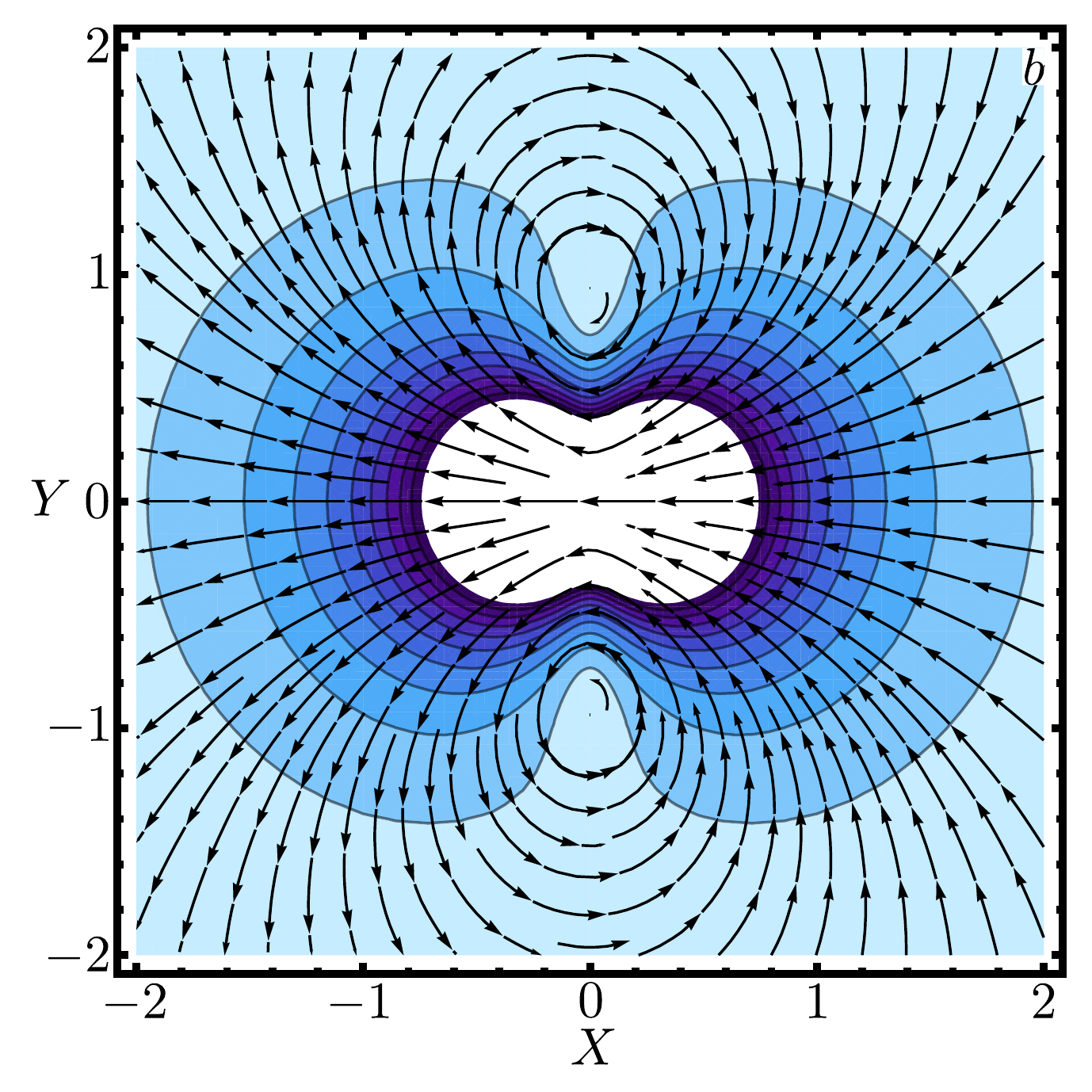}
\includegraphics[height=190pt]{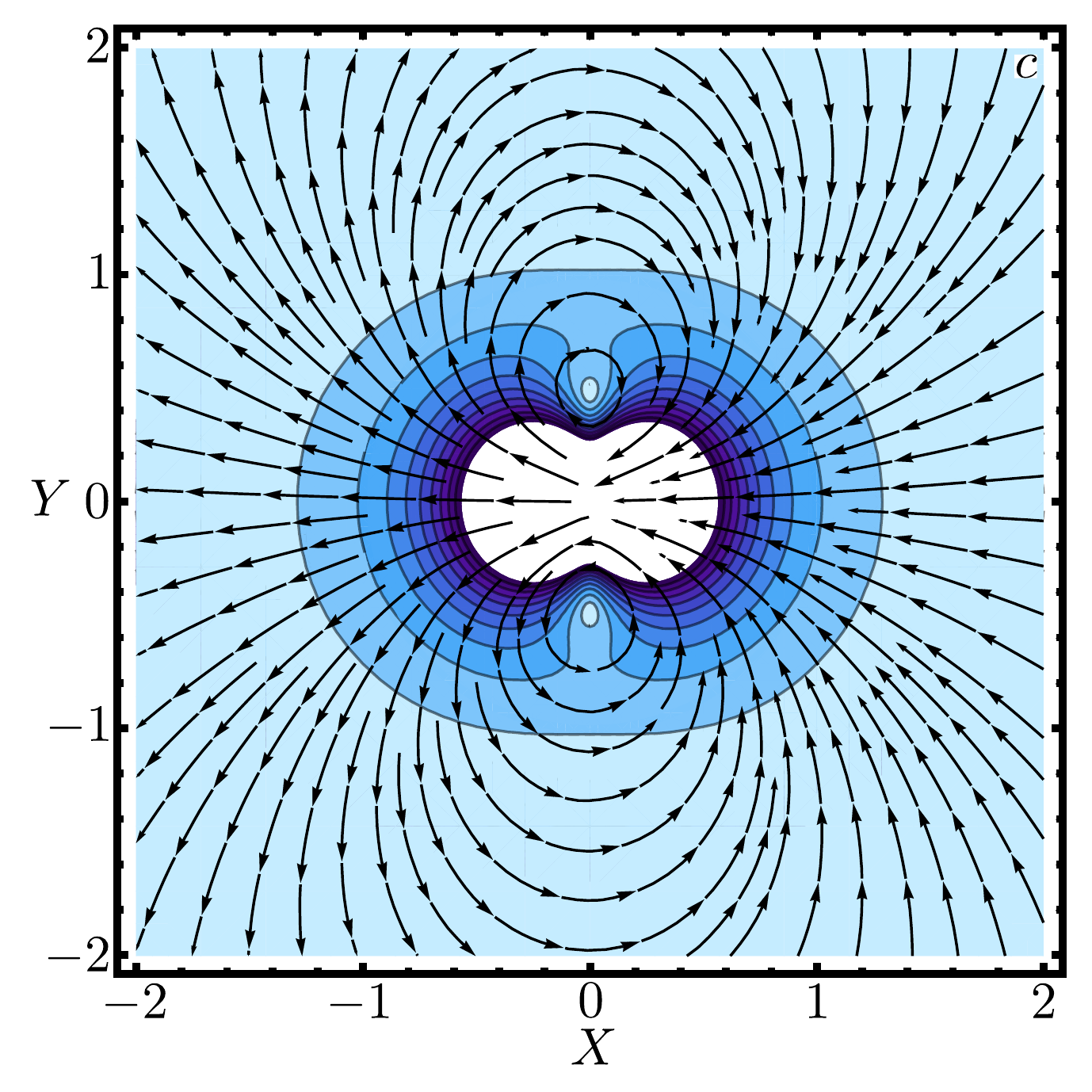}
\includegraphics[height=200pt]{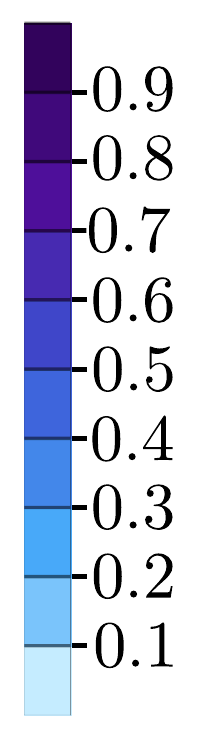}
\caption{{ The difference in flow between the three regularised stokeslets and the true singularity. (a) The difference in the power law regularisation, $(\mathbf{S}_{\epsilon}^{p}-\mathbf{S})\cdot\mathbf{F}$; (b) The difference in the Gaussian regularisation $(\mathbf{S}_{\epsilon}^{g}-\mathbf{S})\cdot\mathbf{F}$; and (c) The difference in the supported regularisation $(\mathbf{S}_{\epsilon}^{s}-\mathbf{S})\cdot\mathbf{F}$. Arrows display the direction of the deviation flow while contours display the strength. In all cases $z=0$, $\mathbf{F} = 6 \pi \mathbf{\hat{x}}$ and all lengths has been scaled by $\epsilon$. }}
\label{fig:deviation}
\end{figure}

 The substitution of a blob for the Dirac delta removes the divergent flow from the resultant singularity. Cortez called the fundamental of these structures the `regularised stokeslet', solution to
\begin{eqnarray}
-\nabla \mathbf{P}_\epsilon + \nabla^2\mathbf{S}_\epsilon +f_\epsilon\mathbf{I} &=&\mathbf{0},\\
\nabla\cdot\mathbf{S}_\epsilon &=& 0,
\end{eqnarray}
 where $\mathbf{S}_\epsilon$ is the regularised Oseen tensor and $\mathbf{P}_\epsilon$ is the regularised pressure with regularization parameter $\epsilon$ and blob $f_{\epsilon}$. The flow from a regularised stokeslet is unique to each regularisation blob, $f_\epsilon(\mathbf{r})$, but  always has a sharp peak at the origin for small $\epsilon$ and a longer tail for big $\epsilon$ (Fig.~\ref{figure_Comparison}). The regularised stokeslet tensors for power-law, $\mathbf{S}_{\epsilon}^{p}(\mathbf{r})$, Gaussian, $\mathbf{S}_{\epsilon}^{g}(\mathbf{r})$, and supported blob, $\mathbf{S}_{\epsilon}^{s}(\mathbf{r})$, are given  respectively by
 \begin{eqnarray}
 \mathbf{S}_{\epsilon}^{p}(\mathbf{r}) &=&\frac{\mathbf{I}(r^2 + 2\epsilon^2) +
\mathbf{rr}}{8\pi r_\epsilon^3}, \label{power} \\
\notag \\
\mathbf{S}_{\epsilon}^{g}(\mathbf{r}) &=& \mathbf{I}\left(\frac{(\epsilon^2 +
2r^2)\text{erf}(r/\epsilon)}{16\pi r^3} -\frac{
2 r\epsilon}{16\pi\sqrt{\pi} r^3} e^{-r^2/\epsilon^2} \right)\nonumber\\
&&+
\mathbf{rr}\left(\frac{3\epsilon}{8\pi\sqrt{\pi}r^4}e^{-r^2/\epsilon^2}
- \frac{3\epsilon^2\text{erf}(r/\epsilon)}{16\pi r^5} +
 \frac{\text{erf}(r/\epsilon)}{8\pi r^3}\right), \label{guass} \\
 \notag \\
 \mathbf{S}_{\epsilon}^{s}(\mathbf{r}) &=&\displaystyle \left\{\begin{array}{c l}
\displaystyle \mathbf{I}\left(\frac{5}{16\pi\epsilon} -
\frac{r^2}{4\pi\epsilon^3} + \frac{9r^4}{112\pi\epsilon^5}\right)
+ \mathbf{rr}\left(\frac{1}{8\pi\epsilon^3} -
\frac{3r^2}{56\pi\epsilon^5}\right), & r<\epsilon \\
\displaystyle \mathbf{I} \left(\frac{1}{8\pi
r} + \frac{\epsilon^2}{56\pi
r^3} \right) + \mathbf{rr} \left(\frac{1}{8\pi
r^3} -\frac{3 \epsilon^{2}}{56 \pi 
r^5} \right) ,&r>\epsilon.
\end{array} \right.  \label{support}
 \end{eqnarray}
 In the above $r_\epsilon=\sqrt{r^2+\epsilon^2}$ and  $\text{erf}(r)=\displaystyle \frac{2}{\sqrt{\pi}}\int_0^r e^{-t^2}dt$ is the error function. The structure of these flows are very distinct {(see Fig.~\ref{fig:deviation})}. The power-law flow decays with negative powers of $r$, while the Gaussian has flows that decay exponentially. The flow from supported blob is different again, growing inside the support while decaying outside it. Different regularisations clearly produce very different flows and so it is important to further understand these flows to obtain insight on the accuracy of regularised stokeslet simulations. 

\section{Taylor expansion of example regularised stokeslets} \label{sec:taylor}

 Flows from regularised stokeslets limit to the stokeslet as $\epsilon\to0$. This is a consequence of the blobs convergence to the Dirac delta as $\epsilon \rightarrow 0$ and offers a natural position to explore the physics of the regularised flows.  A Taylor series expansion of the regularised flow around $\epsilon=0$ should write the flow as a stokeslet and higher-order structures proportional to powers of $\epsilon$. The first non-zero correction determines the convergence of the flow. In this section, we carry out the  expansion for the three example regularised stokeslets, Eqs.~\eqref{power}, \eqref{guass} and \eqref{support}, in order to develop an intuition for flows and the possible variations. These expansions will be performed at fixed $\mathbf{r}$ and will  typically be valid for $\epsilon<{\cal O}(r)$.

\subsection{Power-law blob}

The power-law blob has the slowest decay rate among the three examples. The resultant regularised stokeslet, Eq.~\eqref{power}, is therefore anticipated to have a long tail. This flow can be expanded around $\epsilon=0$ by moving $1/r^3$ out of the fraction and applying the binomial theorem to $[1+(\epsilon/r)^2]^{-3/2}$. This expansion has the form
\begin{eqnarray}
\mathbf{S}_{\epsilon}^{p}(\mathbf{r})&=& \frac{1}{8\pi r^3}\frac{\mathbf{I}(r^2 + 2\epsilon^2) + \mathbf{rr}}{\sqrt{1+(\epsilon/r)^2}^3}= \left(\mathbf{S}(\mathbf{r}) +
\frac{2\mathbf{I}}{8\pi r^3}\epsilon^2\right)\left[1-\frac{3\epsilon^2}{2r^2} +
\frac{15\epsilon^4}{8r^4} - \frac{35\epsilon^6}{16r^6} +
{\cal O}(\epsilon^8)\right]\notag\\
&=& \mathbf{S}(\mathbf{r}) +
\frac{\mathbf{I}r^2-3\mathbf{rr}}{8\pi r^5}\frac{\epsilon^2}{2!} +
\left(-\frac{27\mathbf{I}}{8\pi r^5} +
\frac{45\mathbf{rr}}{8\pi r^7}\right)\frac{\epsilon^4}{4!}
+\left(\frac{25\mathbf{I}}{128\pi r^7}-\frac{35\mathbf{rr}}{128\pi r^9}\right)\epsilon^6 + {\cal O}(\epsilon^8) \notag \\
&=&\mathbf{S}(\mathbf{r})+\frac{\epsilon^2}{2}\mathbf{D}(\mathbf{r})+\ldots \label{power_expand}
\end{eqnarray}
The first term of this expansion is the stokeslet, as expected, while the 
second term is proportional to the
source dipole. Higher-order terms are not singularity solutions. 
Notably, none of the odd terms survive but the even terms go on
forever. The flow for $\epsilon<r$  converges to the stokeslet flow at order $\epsilon^{2}$. The higher-order non-singularity terms
contribute little to the far-field flow, $\epsilon \ll r$, as they are at least $\epsilon^2/r^2$ smaller than the dipole term. However in the limit $r\sim\epsilon$ their contribution can obviously become similar to that of the stokeslet and source dipole. This shows that the regularisation significantly modifies the flows at $r\lesssim\epsilon$ and so $\epsilon$ should be kept much smaller than all other system lengths. Notably the limit $r \ll \epsilon$ produces an isotropic flow of $\mathbf{S}_{\epsilon}^{p}(\mathbf{r}) = \mathbf{I}/(4\pi\epsilon) + {\cal O}(r^2/\epsilon^2)$, in  contrast with the stokeslet which is always anisotropic. This is common to many regularised stokeslets.

\subsection{Gaussian blob}

The Gaussian blob decays exponentially and so should produce a more compact flow. A Taylor series expansion of Gaussian regularised stokeslet, Eq.~\eqref{guass}, clearly demonstrates this, with
\begin{equation}\label{eqn_Gaussian_Expansion}
\mathbf{S}_{\epsilon}^{g}(\mathbf{r})=\mathbf{S}(\mathbf{r}) + \frac{\epsilon^2}{2}\mathbf{D}(\mathbf{r}) + {\cal O}(\epsilon^n),
\end{equation}
for all positive integer $n$. The above form occurs because the error function $\text{erf}(r)$ converges exponentially to $1$ when $r\to\infty$. Exponentially small corrections are therefore needed for the series to converge to the regularised flow, Eq.~\eqref{guass}. Here again the deviation from the stokeslet includes a second order term proportional to the source dipole, but the series expansion has no non-singularity contributions. The Gaussian regularisation is, therefore, a better approximation of the Dirac delta than the power-law blob and could be used for singularity representations in which the stokeslet and source dipole occur in combination \cite{ChwangWu}. In these cases the regularisation parameter, $\epsilon$, takes the role of a physical length, as was proposed in a number of regularised slender-body theories \cite{Bouzarth,Smith}.

\subsection{Compact blob}

The compact support blob has the fastest decay, since it is identically 0 for $r>\epsilon$. The flow outside the supported blob, Eq.~\eqref{support}, only contains terms proportional to $1=\epsilon^{0}$ and $\epsilon^{2}$. A rearrangement of this flow produces
\begin{equation}
\mathbf{S}_{\epsilon}^{s}(\mathbf{r}) = \displaystyle \mathbf{S}(\mathbf{r})+\frac{\epsilon^2}{7} \mathbf{D}(\mathbf{r}) , \quad r>\epsilon.
\end{equation}
Hence the flow is exactly a linear combination of two fundamental singularities, the stokeslet and source dipole. The inside flow cannot be written similarly. Again there is no non-singularity flows present in this outer flow. This suggests that their existence is related to the decay rate of the regularisation. 
Irrespective of the decay, the source dipole has been present in all the flows. This implies that the regularisation of the stokeslet introduces an additional source dipole flow at $\epsilon^{2}$ and higher-order terms depending on the decay of the blob. These trends are explained and generalised in the following sections.

\section{Expansion of a general regularised stokeslet} \label{sec:convolution}
The far-field behaviour of a specific regularised stokeslet
can be identified through a Taylor series. A general
representation of the flow is needed to produce a similar
expansion for a regularised stokeslet with arbitrary blob
$f_\epsilon(\mathbf{r})$. The linearity of Stokes flow
allows the total flow from any blob to be written as
\begin{equation}	
\mathbf{S}_\epsilon(\mathbf{r}) = \mathbf{S}*f_\epsilon(\mathbf{r}) =
 \frac{1}{\epsilon^3}
\iiint \mathbf{S(r - r')}f(\mathbf{r'}/\epsilon)\,d\mathbf{r'}=
\iiint \mathbf{S(r-\epsilon r')}f(\mathbf{r'})\,d\mathbf{r'},
\label{eqn_Integral_Form_RStokeslet}
\end{equation} 
where $f*g(\mathbf{r})=g*f(\mathbf{r}) = \iiint
f(\mathbf{r'})g(\mathbf{r-r'})\,d\mathbf{r'}$ denotes the
convolution between $f$ and $g$. Note that higher-order regularised singularities cannot be created through an equivalent convolution because the derivatives of the stokeslet have non-integrable singularities. Formally the integrand of the convolution representation is differentiable at $\epsilon=0$ and so the integral has a Taylor series of
\begin{equation}
\mathbf{S}_\epsilon(\mathbf{r}) = \mathbf{S}(\mathbf{r}) -
\epsilon\nabla\mathbf{S}(\mathbf{r})\cdot\iiint\mathbf{r'}f(\mathbf{r'})\,
d\mathbf{r'}
+
\frac{\epsilon^2}{2!}\nabla\nabla\mathbf{S}(\mathbf{r}) 
:\iiint\mathbf{r'}\mathbf{r'}f(\mathbf{r'})\,d\mathbf{r'}
+\cdots. \label{general}
\end{equation}
This series represents the flow solely in terms of singularity solutions to Stokes flow. Importantly, the coefficients in this series may not always { be} correct because the expansion can encounter two difficulties: (i) the integrand of the $n^{\rm th}$  term in the expansion,
$\mathbf{r}^n f(\mathbf{r})$, might not decay fast enough
for the integral to exist (specifically $\iiint\mathbf{r'}^n f(\mathbf{r'})d\mathbf{r'}$ diverges if $\mathbf{r'}^n f(\mathbf{r'})$ scales as $r'^{-3}$ or slower) or (ii) the divergent flow at the origin of the stokeslet means the differentiation and integration cannot always be interchanged. These difficulties cause the flow from a general regularised stokeslet to behave
differently to the above expansion beyond a certain power of $\epsilon$. The non-singularity terms in the power-law blob expansion, Eq.~\eqref{power_expand}, reflect this. The error and  validity of the general Taylor expansion  reveal therefore information about the flow induced by regularised stokeslets and are needed to understand the expansion.

\subsection{Error and validity of the general Taylor expansion} 

The error and validity of the general expansion determine to what order in $\epsilon$ the flow from a regularised stokeslet is accurately represented by flow singularities. Hence the identification of this error is needed to understand the flow. In this subsection, we evaluate this error for blobs of the from $f(r)\lesssim r^{-n-5}$ where $\lesssim (\cdot)$ means $\leq C (\cdot)$ for some constant $C$. This condition includes all the examples considered above.

The remainder (error), $E_{n}$, between a function $g(x+\epsilon)$ and its Taylor series, in $\epsilon$, to order $n$ is known to be bounded by 
\begin{equation}\label{eqn_LagrangeRemainder}
|E_n|\leq\frac{\max |g^{(n+1)}(x)|}{(n+1)!}\epsilon^{n+1},
\end{equation}
where $g^{(n+1)}(x)$ is the $n+1^{\rm th}$ derivative of $g(x)$. For the expansion of a stokeslet of the form  $\mathbf{S(r-\epsilon r')}$, this bound is infinite for any $n$ because every derivative of the stokeslet diverges at their origin. This issue can be overcome by isolating the singularity in the stokeslet from the expansion region while not compromising the function's smoothness.

\begin{figure}
\centering
\includegraphics[height=180pt]{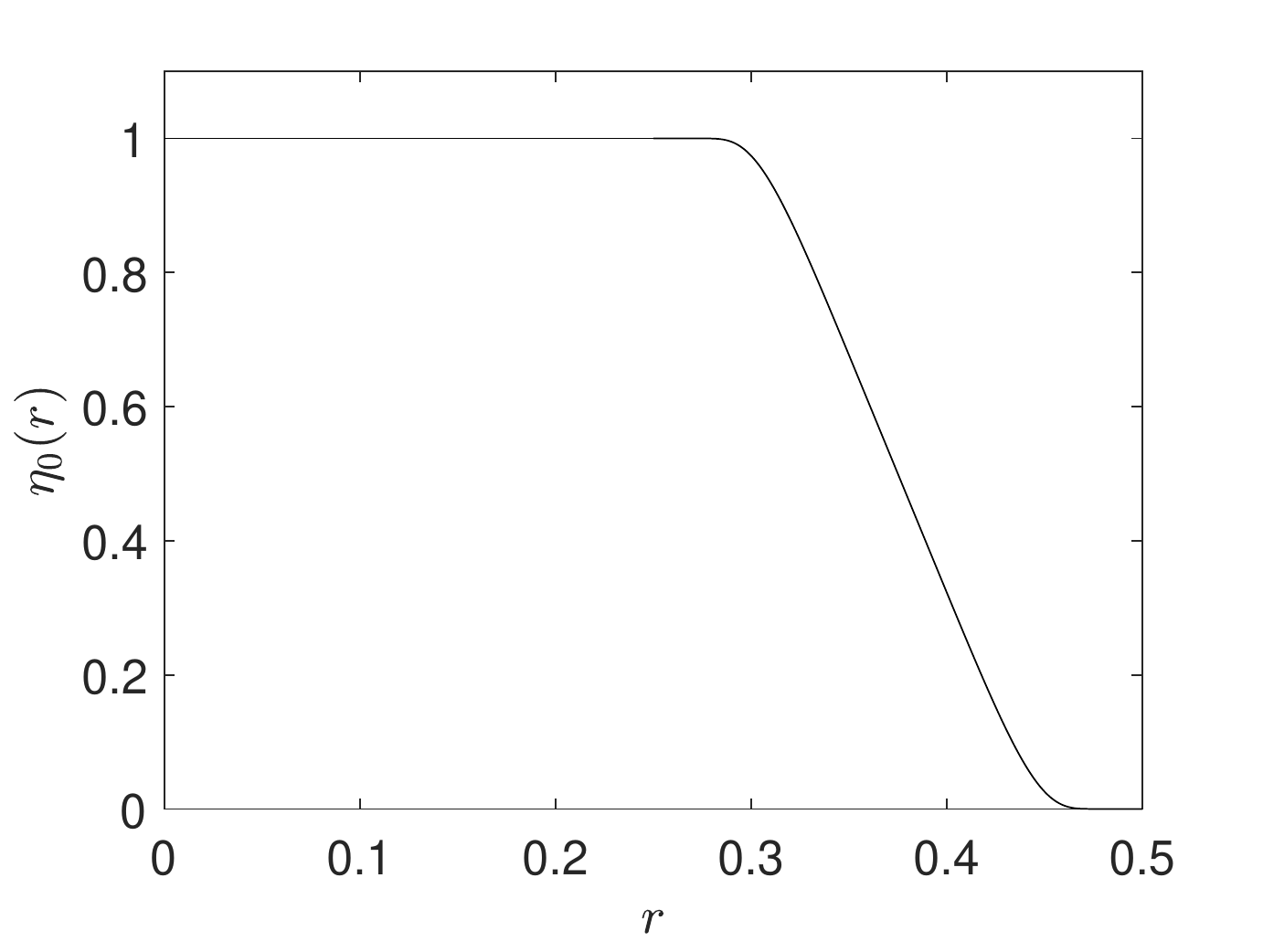}
\caption{Plot of the example cutoff function $\eta_{0}(r)$, Eq.~\eqref{eta}.}
\label{fig_eta}
\end{figure}

The separation of the singularity from smooth parts of the stokeslet can be achieved by multiplying by a cutoff function $\eta(\mathbf{r})$ which has compact support and equals
1 around the origin. This function $\eta(\mathbf{r})$ must smoothly transition from the $\eta(\mathbf{r})=1$ near the origin to the cutoff region. This would not work for the  compact support blob, Eq.~\eqref{support},   because it is not differentiable at $\epsilon$. An example of a $\eta(\mathbf{r})$ function is
\begin{equation}
\eta(\mathbf{r})=\eta_{0}\left(\frac{r}{a}\right)  = \left\{\begin{array}{c c}
1 &\displaystyle  r\leq\frac{a}{4}, \\
\frac{1}{2}\left\{\displaystyle \mbox{erf}\left[\tan\left(\frac{3 \pi}{2} -  \frac{4 \pi r}{a}\right)\right] +1\right\} & \displaystyle  \frac{a}{4}<r< \frac{a}{2},\\
0 & \displaystyle  r\geq \frac{a}{2},
\end{array} \right. \label{eta}
\end{equation} 
where  $\rm erf$ is the error function defined above and $a/2$ is the radius of the support. This example of a cutoff function is plotted in Fig.~\ref{fig_eta}. It smoothly connects the two flat regions and is infinitely differentiable at all $r$. The product of $\eta(\mathbf{r})$ with any other smooth function is also smooth, exactly mimics the behaviour of the function within a ball of radius $r=a/4$ and is supported within a ball of radius $r=a/2$. 

The cutoff function allows the stokeslet to be smoothly split into two parts, $\mathbf{A(r)}=\eta(\mathbf{r})
\mathbf{S(r)}$ and $\mathbf{B(r)=S(r)-A(r)}$. The
first part, $\mathbf{A(r)}$, vanishes when $r>a/2$ and
captures the singularity at $r=0$ while the second part,
$\mathbf{B(r)}$, is smooth, agrees with the stokeslet
when $r>a/2$ and vanishes when $r<a/4$. By definition, $\mathbf{S(r) = A(r) + B(r)}$ and so the convolution integral can be written as
\begin{equation}\label{eqn_splitting_integral}
\iiint
\mathbf{S(r-\epsilon r')}f(\mathbf{r')}\,d\mathbf{r'}=\iiint
\mathbf{A(r-\epsilon r')}f(\mathbf{r')}\,d\mathbf{r'}+\iiint
\mathbf{B(r-\epsilon r')}f(\mathbf{r')}\,d\mathbf{r'}.
\end{equation}
The first of the integrals on the right hand side contains the singularity and so does not have a convergent Taylor expansion. The second integral however is smooth and has bounded derivatives and so its expansion around $\epsilon=0$ exists. This expansion is expected to form the proposed Taylor series structure in Eq.~\eqref{general}. This requires $\mathbf{B}(\mathbf{r})$ and $\mathbf{S}(\mathbf{r})$ to have the same Taylor expansion around the point of interest, $\mathbf{r}$. Since by definition $\mathbf{B}(\mathbf{r}) =  \mathbf{S}(\mathbf{r})$ when $|\mathbf{r}|=r>a/2$, this condition implies that $a = k r$, where $k$ is an arbitrary constant between 0 and 2. We use $a$ throughout the arguments and substitute in the relation, $a = k r$, at the ends to avoid any confusion between actual locations and the radius of support of $\eta(\mathbf{r})$.  

The error on the general Taylor expansion, Eq.~\eqref{general}, can be found from Eq.~\eqref{eqn_splitting_integral} by determining the size of the singularity term and the remainder from the Taylor expansion of the smoothed term. The size of the singularity term is bounded by
\begin{eqnarray}
\left|\iiint \mathbf{A(r-\epsilon r')}f(\mathbf{r')}\,d\mathbf{r'}\right|
&=&\epsilon^{-3}\left|\iiint
\mathbf{A(r-r')}f(\mathbf{r'/\epsilon)}\,d\mathbf{r'}\right| \nonumber\\
&\lesssim& \epsilon^{-3}\iiint \left|\mathbf{A(r-r')}\right|\sqrt{1 +
|\mathbf{r'}/\epsilon|^2}^{-n-4}\,d\mathbf{r'}\nonumber\\
&\leq&\epsilon^{n+1}\iiint \left|\mathbf{A(r-r')}\right|
r'^{-n-4}\,d\mathbf{r'}\nonumber\\
&\lesssim&\frac{\epsilon^{n+1}}{(2r-a)^{n+4}}\iiint
\left|\mathbf{A(r-r'})\right|\,d\mathbf{r'},
\end{eqnarray}
where we have recognised that blobs with $f(r)\lesssim r^{-n-5}$ satisfy the weaker condition of $f(r)\lesssim r^{-n-4}$. In the last inequality we used that $\mathbf{A(r-r')}$, as a function of $\mathbf{r'}$, is non-zero in a ball of radius $a/2$ around $r$ and so $r'$ is smallest when $r' = r -a/2$ since $a/2 < r$.   The remaining integral grows like
$\int_0^a 1/r'\times r'^2 dr' \sim a^{2}$ since the stokeslet, $\mathbf{S(r)}$, decays
like $1/r$. Hence the size of the singularity term is bounded by
\begin{equation}\label{eqn_estimate_A}
\left|\iiint \mathbf{A(r-\epsilon r')}f(\mathbf{r')}\,d\mathbf{r'}\right|
\lesssim\frac{\epsilon^{n+1}}{(2r-a)^{n+4}}\iiint
|\mathbf{A(r-r'})|\,d\mathbf{r'} \lesssim\frac{\epsilon^{n+1} a^{2}}{(2r-a)^{n+4}} \sim \frac{\epsilon^{n+1}}{r^{n+2}}  ,
\end{equation}
where the constant in the inequality depends only on the blob
and $n$ and we have used $a = k r$. This bound holds globally and so provides an accurate estimate of the size of the singularity term.  

The error produced in the expansion of second term can be determined through the remainder produced by expanding $\mathbf{B(r - \epsilon r')}$. Since $\mathbf{B(r)}$ is always well behaved, the remainder from expanding $\mathbf{B(r - \epsilon r')}$ is bounded by
\begin{eqnarray}\label{eqn_Expansion_B}
\left|\mathbf{B(r-\epsilon r')}-\left(\mathbf{S(r)-\epsilon\nabla S(r)\cdot r'+
}\ldots+\frac{(-\epsilon)^n }{n!}\nabla^n\mathbf{S(r)}\cdot
\underbrace{\mathbf{r'\ldots
r'}}_{\text{n copies}}\right)\right|\nonumber\\
\leq\frac{\epsilon^{n+1}\mbox{ }  r'^{n+1}}{(n+1)!}\max\left|\nabla^{n+1}\mathbf{B(r)}
\right|,
\end{eqnarray}
where $\cdot$ denotes the k-fold contraction. In the above we used that $\mathbf{B(r)} = \mathbf{S(r)}$ for $a/2 < r$ and so the Taylor coefficients of $\mathbf{B(r)}$ can be replaced by the Taylor coefficients of $\mathbf{S(r)}$. The size of the $n+1^{\rm th}$ derivative of $\mathbf{B(r)}$ therefore determines the remainder. The gradient operator, $\nabla$, scales as 1/length, and so the $n$th derivative of $\mathbf{B(r)}=\mathbf{S(r)}(1-\eta(\mathbf{r}))$ scales with (1/length)$^n$ times the original function. This 1/length factor must be less than or equal to $4/a$ because $\mathbf{B(r)}$ is 0 inside the ball $r<a/4$. Hence the size of the $n+1^{\rm th}$ derivative is bounded by
\begin{equation}
|\nabla^{n+1}\mathbf{B(r)}|\lesssim a^{-(n+2)}.
\end{equation}
 This makes the remainder in the Taylor expansion
\begin{equation}
\frac{\epsilon^{n+1}\mbox{ }  r'^{n+1}}{(n+1)!}\max\left|\nabla^{n+1}\mathbf{B(r)}
\right| \lesssim \frac{\epsilon^{n+1}\mbox{ }  r'^{n+1}}{a^{n+2}}.
\end{equation}
Multiplying both both sides of \eqref{eqn_Expansion_B} by the absolute value of the blob $|f(\mathbf{r'})|$ and integrating over $\mathbf{r'}$, then indicates that
\begin{eqnarray}\label{eqn_expansion_BIntegral}
&&\left|\iiint \mathbf{B(r-\epsilon r')}f(\mathbf{r')}\,d\mathbf{r'} -\left(
 \mathbf{S(r) -
\epsilon\nabla S(r)\cdot\iiint
 r'}f(\mathbf{r'})\,d\mathbf{r'}\right.\right.
\nonumber\\
&& \left.\left.+\cdots+ \frac{(-\epsilon)^n}{n!}\nabla^n\mathbf{S(r)}\cdot\iiint
\underbrace{\mathbf{r'\ldots
r'}}_{\text{n copies}}f(\mathbf{r'})\,d\mathbf{r'}\right)\right|\lesssim
\frac{\epsilon^{n+1}}{a^{n+2}} \sim \frac{\epsilon^{n+1}}{r^{n+2}},
\end{eqnarray}
where we have used that the radius of support $a$ must be proportional to the distance $r$ for the series to capture the desired form, and the integral $\iiint r'^{n+1}|f(\mathbf{r'})|\,d\mathbf{r'}$ is guaranteed to exist due to the decay condition $f(r)\lesssim r^{-n-5}$. This result demonstrates that the error on the expansion scales as $\epsilon^{n+1}/r^{n+2}$, identically to that found for the singularity contribution, Eq.~\eqref{eqn_estimate_A}.
 
The two error analyses can be combined to determine the full error by adding up the respective terms. As a result, the expansion of the generalised
regularised stokeslet is
\begin{eqnarray}\label{eqn_expansion_regulariedstokeslet}
\mathbf{S^\epsilon(r)}&=&\iiint \mathbf{S(r-\epsilon r')}f(\mathbf{r')}\,d\mathbf{r'} =
 \mathbf{S(r) -
\epsilon\nabla S(r)\cdot\iiint
 r'}f(\mathbf{r'})\,d\mathbf{r'}
\nonumber\\
&& +\cdots+  \frac{(-\epsilon)^n}{n!}\nabla^n\mathbf{S(r)}\cdot\iiint \underbrace{\mathbf{r'\ldots
r'}}_{\text{n copies}}f(\mathbf{r'})\,d\mathbf{r'}+E(\mathbf{r},\epsilon),
\end{eqnarray}
where $E(\mathbf{r},\epsilon)$ is the error on the expansion,  bounded by
\begin{equation}\label{eqn_final_error}
|E(\mathbf{r},\epsilon)|\leq C\frac{\epsilon^{n+1}}{r^{n+2}},
\end{equation}
for all $r \neq 0$ and $\epsilon>0$. In the above $C$ is a constant that depends only on the chosen blob and on the value of  $n$.

Physically this error tells us to what order the flow from a regularised stokeslet can be described using singularity solutions to the Stokes equations. The flow generated from regularisation blobs decaying as $f(r)\sim r^{-n-5}$ is composed of singularity solutions to order $n$, beyond which non-singularity terms may arise.  The power-law regularised stokeslet, Eq.~\eqref{power}, has $n = 2$ and so the flow can be written in terms of singularity solutions until $\epsilon^2$, beyond which non-singularity solutions become necessary. This
general result is unable to tell us the form of these higher
non-singularity components as they do not usually satisfy the
homogeneous ($\mathbf{f}$ = 0) Stokes equations. The Gaussian
regularised stokeslet, Eq.~\eqref{guass}, can always be written in
terms of singularity solutions because its blob decays
exponentially. Finally, it is worth noting that the bound obtained in  Eq.~\eqref{eqn_final_error} is global while the Taylor series for a regularised stokeslet usually converge for a specific range of $\epsilon$. The power-law blob has a Taylor series that converges to the regularised stokeslet when
$\epsilon<r$ while the Gaussian blob has a finite Taylor series
that does not sum up to the regularised stokeslet due to missing exponentially small contributions.
 
\subsection{Far-field flow of a general regularised stokeslet}

The generalised series shows that the flow far from the regularised stokeslet is well approximated by an infinite number of singularity solutions weighted by powers of $\epsilon$ when the regularisation blob decays rapidly. These higher-order singularities all vanish as $\epsilon \to 0$ leaving the stokeslet flow. Beyond the stokeslet, the largest contribution comes from the force dipole term that is in general proportional to $\epsilon$. Hence the flow from a general regularised stokeslet converges to the stokeslet as $\epsilon\to 0$ at a rate $\epsilon$. This rate of convergence can be improved if the blob $\mathbf{f(r)}$ has additional symmetries because integrals of the form $\displaystyle \iiint \mathbf{r'}^n f(\mathbf{r'}) \,d\mathbf{r'}$ may disappear for certain $n$. Arbitrary blobs, without any symmetries, will always induce a force dipole flow and higher-order singularities. 

However, if the blob has three distinct reflection symmetries, the singularities proportional to odd powers of $\epsilon$ become zero.  Ellipsoidal blobs have such reflection symmetries. The far-field flow in this case is composed of an infinite number of flow structures with even powers of $\epsilon$. This improves the convergence of the far flow to the stokeslet to $\epsilon^{2}$. Spherically symmetric blobs simplify this series further as the volume integrals of $\mathbf{r}^{2n}f(\mathbf{r})$ must all be isotropic. The higher singularities are therefore of the from $\nabla^{2 n} \mathbf{S}(\mathbf{r})$, where $n$ is a positive integer. These terms are zero for $n \geq 2$ since Stokes flow is bi-harmonic ($\nabla^{4} \mathbf{S}(\mathbf{r}) = \mathbf{0}$ for $\mathbf{r}\neq \mathbf{0}$). The general expansion for spherically symmetric blobs is therefore of the form
\begin{equation}\label{eqn_Isotropic_Expansion}
\mathbf{S}_\epsilon(\mathbf{r}) = \mathbf{S}(\mathbf{r})
+\frac{4\pi \epsilon^2}{3}\mathbf{D}(\mathbf{r})\int_{0}^{\infty} r^4f(r)\,dr + {\cal O}(\epsilon^{n}), 
\end{equation}
 where $n\geq3$, and we have used Eq.~\eqref{dipole}. This explains the presence of only a stokelet and source dipole term in the expansions of the Gaussian and compact support examples. The breakdown of the general expansion at $\epsilon^{2}$ provides the power-law distribution with higher terms. Notably, the source dipole term cannot be eliminated when $f(r)$ is non-negative and so is common to all typical blobs.

\section{Spherically symmetric regularisations} \label{sec:sphere}
Regularisation blobs with spherical symmetry are by far the most common structures used in numerical simulations \cite{Cortez2001,Cortez2005,Smith2018,Godinez2015, Ishimoto2018,Montenegro-Johnson2015, Montenegro-Johnson2018,Olson2013,Montenegro-Johnson2012, Smith, HoaNguyen2014, Montenegro-Johnson2016, Cortez2012, Rodenborn2013a, Bouzarth, Ainley2008,Cortez2015}. The flow generated from these regularisation merits  particular attention. In this case the force distribution has similarities to the surface force distribution of a translating sphere. This allows the flow to be significantly simplified using the different representations for the flow around a sphere. In the  next subsections, we  derive and use the  flow representation for general spherically symmetric regularised stokeslet.

\subsection{Sphere representation of spherically symmetric regularised stokeslets}

 The complete flow from a spherically symmetric stokeslet can be written in a compact form by drawing a parallel between a translating sphere and the regularisation blob. A translating sphere in Stokes flow has a uniform surface traction over its surface. Specifically for a sphere of radius $a$, translating with velocity $\mathbf{U}$, the surface traction on the fluid from the sphere is ${3 }\mathbf{U}/{2 a}$ everywhere on the surface. Similarly the force, from a spherically symmetric blob, is uniform over every sphere centred at the origin $f(\mathbf{r}) = f(r)$.  The flow from a spherically symmetric regularised stokeslet is, therefore, equivalent to a superposition of the flow from many translating spheres,  {i.e.}~we can write
\begin{equation}
\mathbf{S}_\epsilon(\mathbf{r}) = \int_{0}^{\infty} \left( \iint \mathbf{S(r-r')}r'^{2}\,d\boldsymbol{\Omega}' \right) f_{\epsilon}(r') \,dr' = \frac{2 }{3} \int_{0}^{\infty} \mathbf{u}_{\mbox{s}}(r',\mathbf{r}) f_{\epsilon}(r') r' \,dr',
\end{equation}
where $\boldsymbol{\Omega}$ is a spherical solid angle and $\mathbf{u}_{\mbox{s}}(a,\mathbf{r})$ is the flow at $\mathbf{r}$ around a translating sphere of radius $a$ with unit velocity, $\mathbf{I}$. The factor of $2/3$ arises produce the correct traction on the fluid. 

The flow outside a translating sphere of radius $a$ can classically be represented exactly by the sum of a stokeslet and a source dipole centred at the origin,
\begin{equation}
\mathbf{u}(\mathbf{r}) = 6 \pi a \mathbf{U}\cdot\left[\mathbf{S(r)} + \frac{a^2}{3}\mathbf{D(r)}\right],
\end{equation}
 while the flow inside is trivially  constant, $\mathbf{u}(\mathbf{r}) =  \mathbf{U}$ \cite{KimKarrila}, where $\mathbf{U}$ is the velocity of the sphere. The uniqueness of Stokes flow then shows 
\begin{equation}
\frac{2 a}{3} \mathbf{u}_{\mbox{s}}(a,\mathbf{r}) = \iint
\mathbf{S}(\mathbf{r-r'})a^{2}\,d\boldsymbol{\Omega}'=\left\{ \begin{array}{c c}
4 \pi a^2\left[\mathbf{S(r)} + \frac{a^2}{3}\mathbf{D(r)}\right] &  |\mathbf{r}|\geq a, \\
{2a}\mathbf{I}/{3} & |\mathbf{r}|\leq a.
\end{array} \right. 
\end{equation}
The above integral is the single-layer boundary integral representation for flow around a translating sphere \cite{Pozrikidis1992}. Single-layer boundary integrals connect the inner and outer flows around objects undergoing rigid body motion.

The flow from a spherically symmetric regularised stokeslet is therefore
\begin{eqnarray} \label{eqn_Sphere_Model}
\mathbf{S^\epsilon}(\mathbf{r})&=&\int_0^{r}\left(\iint
\mathbf{S}(\mathbf{r-r'})r'^{2}\,d\boldsymbol{\Omega}'\right)f_\epsilon(r')\,dr' +
\int_{r}^\infty \left(\iint
\mathbf{S}(\mathbf{r-r'})r'^{2}\,d\boldsymbol{\Omega}'\right)f_\epsilon(r')\,dr'\nonumber\\
&=&\mathbf{S(r)}\int_0^{r/\epsilon}4\pi
r^2f(r)\,dr + \epsilon^2
\mathbf{D(r)}\int_0^{r/\epsilon}\frac{4\pi}{3}
r^4f(r)\,dr +
\frac{2\mathbf{I}}{3\epsilon}\int_{r/\epsilon}^\infty
r f(r)\,dr.
\end{eqnarray}
This representation is valid for all $\mathbf{r}$ and demonstrates that the flow from an arbitrary spherically symmetric regularised stokeslet can always { be} constructed through a combination of stokeslet, source dipole, and isotropic component multiplied by functions of $r$. This is an exact result. The form shows that compact supported regularised stokeslets identically become stokeslets and source dipoles outside their supported region, as seen in the example above. This demonstrates that compact supported distributions are very suitable for singularity representations \cite{ChwangWu}. 
Interestingly, our results mean that the distribution is not uniquely determined by the outer flow as regularisations with the same $\displaystyle \int_{0}^{\infty} f(r)r^4\,dr$ produces the same outer flow. 
\subsection{Spherical far-field}

The far-field flow structure, beyond the stokeslet and source dipoles, can be determined from the sphere representation, Eq.~\eqref{eqn_Sphere_Model}. These structures satisfy
\begin{equation}
\mathbf{S^\epsilon}_{far}(\mathbf{r}) =-\mathbf{S(r)}\int_{|\mathbf{r}|/\epsilon}^\infty4\pi
r^2f(r)\,dr - \epsilon^2
\mathbf{D(r)}\int_{|\mathbf{r}|/\epsilon}^\infty\frac{4\pi}{3}
r^4f(r)\,dr +
\frac{2\mathbf{I}}{3\epsilon}\int_{|\mathbf{r}|/\epsilon}^\infty
r f(r)\,dr,
\end{equation}
where we have subtracted the total stokeslet and source dipole contributions from the sphere representations.
These far-field structures are amenable to regularisation distributions that have a Laurent expansion of the form
\begin{equation}
f(r)=\frac{c_{-6}}{r^6}+\frac{c_{-7}}{r^7}+\ldots
\end{equation}
In this case the flow from the $1/r^n$ component becomes
\begin{equation}
\mathbf{S^\epsilon}_{far}(\mathbf{r}) = \epsilon^{n-3}\left[\frac{4-n}{(n-2)(n-3)(n-5))}\frac{\mathbf{I}}{r^{n-2}}
+ \frac{\mathbf{rr}}{(n-3)(n-5)r^n}\right]. \label{eqn_Higher_Order_Flow}
\end{equation}

\begin{figure}
\centering
\includegraphics[height=200pt]{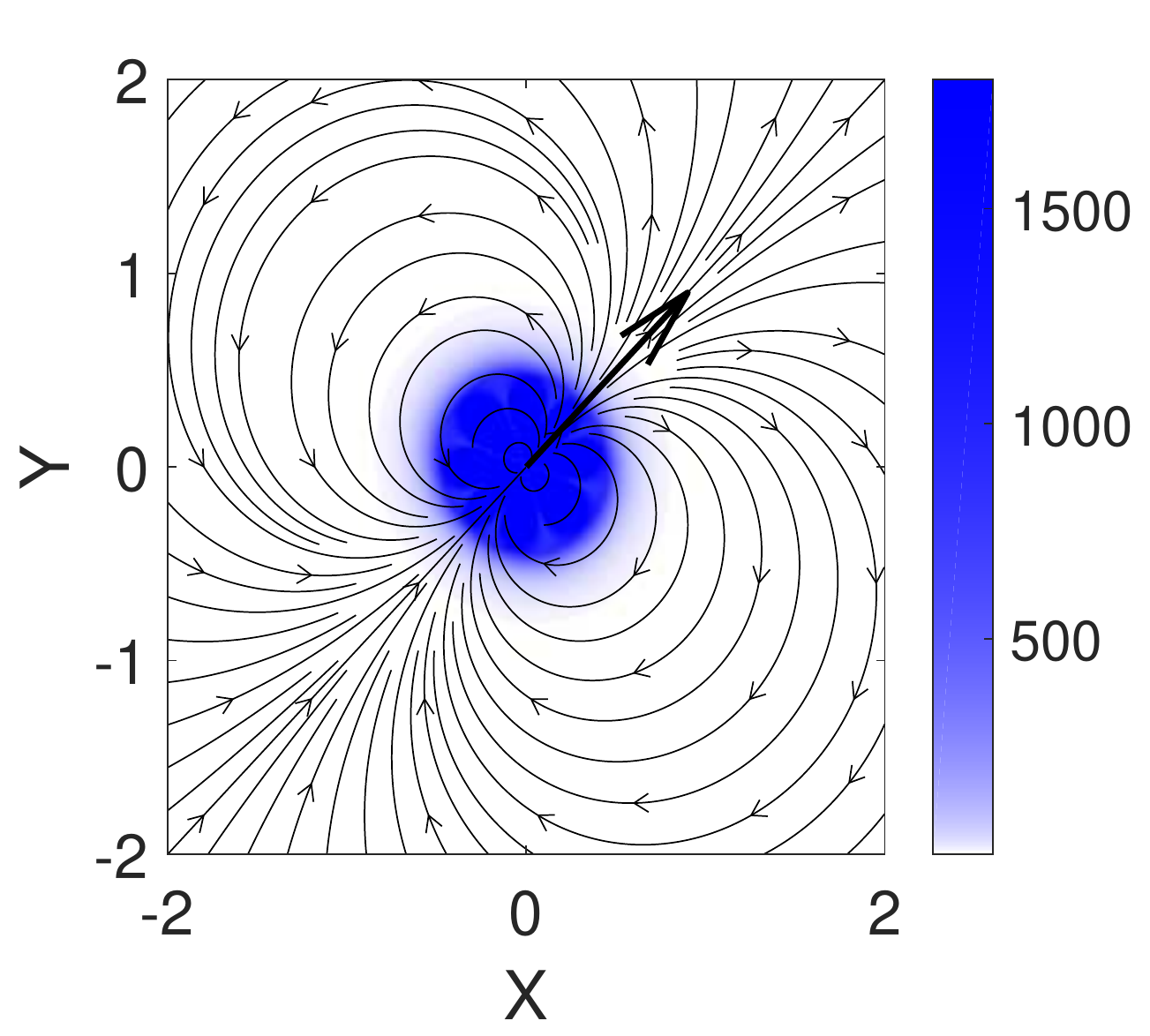}
\caption{Streamlines (lines) and flow strength (density plot) of the $n=7$ flow, $-3\mathbf{I}/5r^5 +\mathbf{xx}/r^7$, with strength (1,1). { The units for $X$ and $Y$ are arbitrary.}}\label{figure_Higher_Order_Flow}
\end{figure}

 This results enables the Taylor expansion of the regularised stokeslet to be immeadiately written down if the Laurent expansion of $f(r)$ is known. For example, the Laurent expansion of the power-law distribution, Eq.~\eqref{power}, is
 \begin{equation}
f_{\epsilon}^{p}(r) \approx \frac{15 \epsilon^{4}}{8 \pi r^{7}} - \frac{105 \epsilon^{6}}{16 \pi r^{9}} + \frac{945 \epsilon^{8}}{64 \pi r^{11}} + {\cal O}\left(\frac{1}{r^{12}}\right),
\end{equation}
and so produces the higher-order terms found in its Taylor expansion, Eq.~\eqref{power_expand}.
The $n=3$ and 5 terms of this expansion are proportional to the stokeslet and the source dipole respectively, however the $n\geq7$ terms do not produce singularity solutions. Physically these flows represent solutions to the Stokes equations with the body force $\mathbf{f}=\mathbf{I}/r^n$ (see Fig.~\ref{figure_Higher_Order_Flow} in the case $n=7$). 
The higher-order flow corrections, Eq.~\eqref{eqn_Higher_Order_Flow}, provide an explicit upper bound on the error in the expansion of an arbitrary spherically symmetric blob, Eq.~\eqref{eqn_Isotropic_Expansion}. The general expansion now becomes 
\begin{equation}
\mathbf{S}_\epsilon(\mathbf{r}) = \mathbf{S}(\mathbf{r})
+\frac{4\pi \epsilon^2}{3}\mathbf{D}(\mathbf{r})\int_{0}^{\infty} r^4f(r)\,dr +
\frac{1}{r}{\cal O}\left(\left(\frac{\epsilon}{r}\right)^{n-3}\right),
\end{equation}
when $f(r) = {\cal O}(1/r^n)$ for $n>5$. The corrections, beyond the source dipole, are always of order $(\epsilon/r)^{n-3}$ smaller than the stokeslet term, $\mathbf{S}(\mathbf{r})$, in such an expansion. These corrections are negligible in the far-field but of course can become significant as the near-field is approached, $r \lesssim \epsilon$.

\subsection{Spherical near field}

In the region very close to the centre of the regularisation, $r\ll\epsilon$, the last term of the spherical representation, Eq.~\eqref{eqn_Sphere_Model}, dominates and the flow reduces to
\begin{equation}
\mathbf{S}_\epsilon(\mathbf{r}) \approx\frac{2\mathbf{I}}{3\epsilon}\int_{0}^{\infty}
rf(r)\,dr.
\end{equation} 
The flow in near centre of all spherically symmetric regularised stokeslets is therefore isotropic, akin the behaviour seen in the power-law example. This isotropy is in contrast to the inherent anisotropy of Stokes flow, as demonstrated by the stokeslet. This difference could have significant implications for numerical simulations. This inner region needs to be kept small, relative to all other lengths, to retain the correct interactions between constituent objects in the flow. The local contribution from the inner region to the surface integrals in the boundary integral representations and the line integrals in slender-body theories also need to be controlled. The respective size of these contributions can be estimated by the strength of the flow in the isotropic component, $2/3\epsilon$, multiplied by the region over which it is significant, $\epsilon^{2}$ in the case of a surface integral and $\epsilon$ in the case of a line integral. The local contribution of the inner region to a surface integral is of order $2 \epsilon/3$ and to a line integral is of $2/3$. The reduction of $\epsilon$ therefore reduces linearly the inner region contribution to any surface integrals and so improves the accuracy of boundary integrals. The contribution to a line integral is independent of $\epsilon$, from this simple scaling argument, and so is always present. These line integrals are typically used in regularised slender-body theories \cite{Smith}, in which { the} regularisation parameter, $\epsilon$, is used to approximate the thickness of a filament. The flow around a slender-body is known to have contributions from an inner cylindrical region  \cite{1976,Koens2018}. The persistent presence of the inner region could benefit these models provided the inner region contribution describes the flow around a cylinder.

\section{Improved accuracy regularised stokeslets} \label{sec:good}

In the previous sections we explored how  geometry, decay rate and blob structure affect the flow produced from a regularised stokeslet. This analysis can then be used to design regularisations that rapidly produce the stokeslet flow. These regularisations should ideally decay at faster than any polynomial, to avoid contributions from body forces, and should be spherically symmetric, to prevent higher-order flow contributions. From  Eq.~\eqref{eqn_Isotropic_Expansion} we also see that the $\epsilon^{2}$ contribution to the far-field flow can also be removed provided that
\begin{equation}
\int_{0}^{\infty} r^4 f(r) \,dr = 0. \label{0dip}
\end{equation}
{ Note that this last condition can only be satisfied if $f(r)$ has negative regions within it. These negative regions are needed to compensate for the spreading out the point force.} Regularisations that satisfy all these additional conditions will converge to the flow from a stokeslet at faster than any polynomial in $\epsilon$ and so have very high accuracy. Here we provide examples of a compact supported regularisation and exponential regulation which satisfies these conditions. We also discuss how the accuracy of power-law regularisations could be improved, though such regularisations will always converge to the stokeslet flow as $\epsilon^{n}$.

\subsection{An improved compact supported regularisation}
\begin{figure}
\centering
\includegraphics[width=\textwidth]{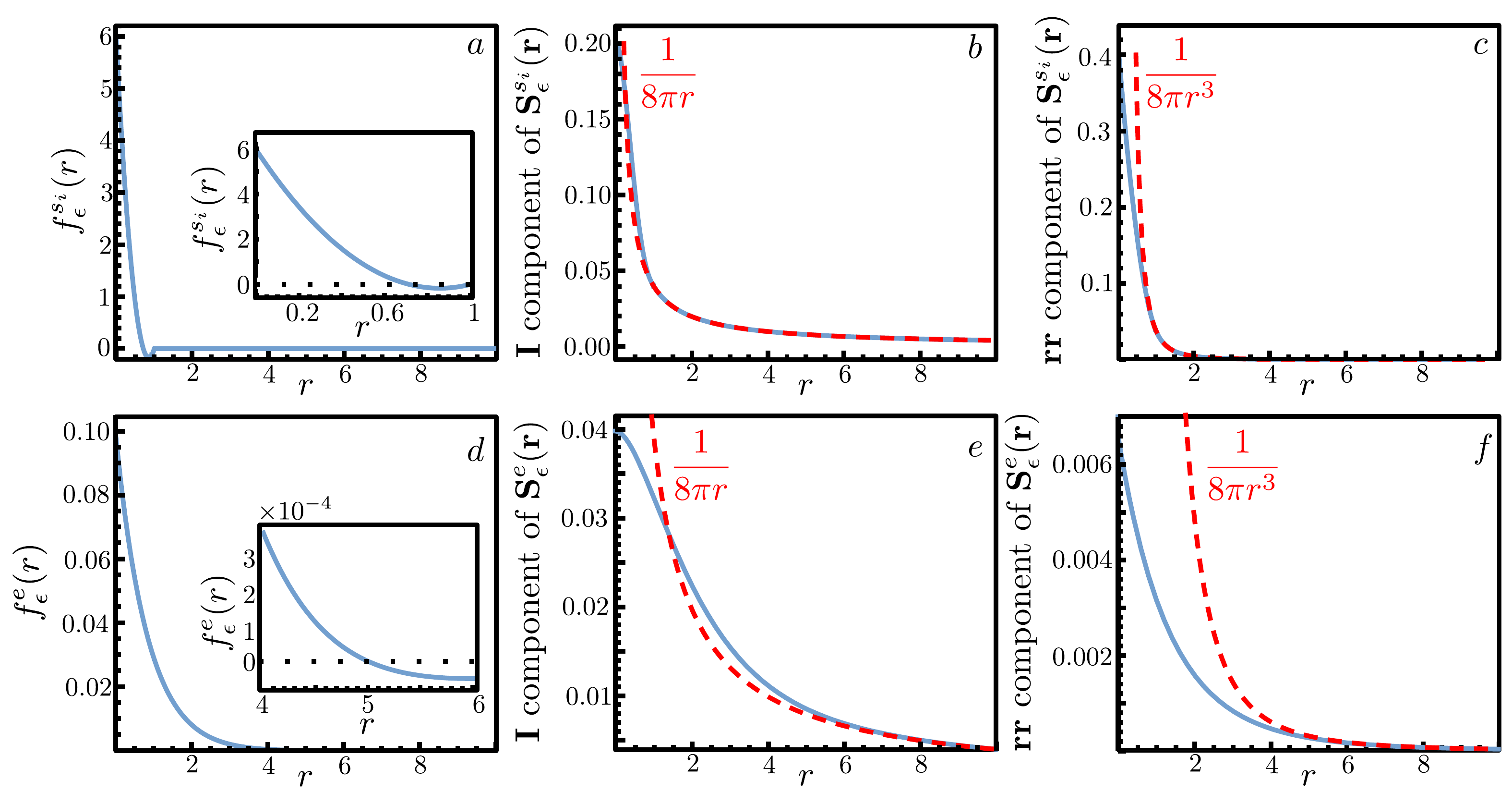}
\caption{The regularisations and flows from the improved-accuracy regularised stokeslets. (a-c): compact supported regularisation; (d-f):  exponential regularisation. Panels (a) and (d) show the improved regularisations. { The insets show the transition where the blobs become negative}; panels (b) and (e) plot the strength of the $\mathbf{I}$ term in the flow; panels (c) and (f) show the strength of the $\mathbf{r}\mathbf{r}$ term in the flow. The red dashed lines are the flow from a stokeslet. { All lengths have been scaled by $\epsilon$.}}
\label{fig:improved_blobs}
\end{figure}

 Compact supported blobs produce the fastest convergence to stokeslets in the far-field flow. The combination of two different supported blobs allows this flow to be further improved by satisfying Eq.~\eqref{0dip} and the normalisation condition. For example, the regularisation blob
 \begin{equation}
 f_{\epsilon}^{s_{i}}(r) = \left\{\begin{array}{c c}
\displaystyle \frac{15}{4\pi\epsilon^5}\left(7r -5 \epsilon\right)\left(r -\epsilon\right), & r<\epsilon \\
0 ,& r> \epsilon
\end{array} \right. , \label{good_fsup}
\end{equation}  
 is constructed by combining a linear, $1-r/\epsilon$, and quadratic regularisation, $1-(r/\epsilon)^2$. In the above $f_{\epsilon}^{s_{i}}(r)$ means the improved supported regularisation. The flow tensor for this regularisation is given by
 \begin{equation}
 \mathbf{S}_{\epsilon}^{s_{i}}(\mathbf{r}) =\displaystyle \left\{\begin{array}{c l}
\displaystyle \frac{1}{8 \pi \epsilon^{5}} \left[\mathbf{I}\left(5 \epsilon^{4} -20 r^{2} \epsilon^{2} +25 r^{3} \epsilon - 9 r^{4} \right)
+ \mathbf{rr}\left(6 r^{2} -15 r \epsilon + 10 \epsilon^{2} \right) \right], & r<\epsilon \\
 \mathbf{S}(\mathbf{r})  ,& r>\epsilon
\end{array} \right. , \label{good_ssup}
 \end{equation}
  as determined by Eq.~\eqref{eqn_Sphere_Model}.  This new regularisation blob and flow are illustrated in Fig.~\eqref{fig:improved_blobs}a-c. This flow does not diverge at $r=0$ and is identical to the stokeslet flow outside the ball of support. { Similar to Eq.~\eqref{support}, derivatives of this flow tensor can be discontinuous. This regularisation is therefore suitable for numerical computations involving the stokelet alone (i.e. boundary integrals and some slender body representations)  but is less suited to singularity representations in which higher-order singularity solutions are needed. }

\subsection{An improved exponential regularisation}
Though a compact supported blob has the fastest convergence, its derivatives can be discontinuous at $r=\epsilon$. Hence an exponential regularisation, which is continuous everywhere, might be preferable. This can again be constructed by combining two exponential regularisations. For example, the exponential regularisation blob
\begin{equation}
 f_{\epsilon}^{e}(r) = \frac{5 \epsilon - r}{16 \pi \epsilon^{2}} e^{-r/\epsilon}, \label{good_fexp}
\end{equation}
is constructed from $e^{-r/\epsilon}$ and $r e^{-r/\epsilon}$ and produces the flow tensor
\begin{equation}
\mathbf{S}_{\epsilon}^{e}(\mathbf{r}) =  \frac{1}{16 \pi \epsilon^{2}} \left[\frac{\mathbf{I}}{r}\left(r^{2} e^{-r/\epsilon} + 2 \epsilon^{2} (1-e^{-r/\epsilon}\right) - \frac{\mathbf{r}\mathbf{r}}{r^{3}}\left(r^{2} e^{-r/\epsilon}+ 2 r \epsilon e^{-r/\epsilon} - 2 \epsilon^{2} (1-e^{-r/\epsilon}\right)\right]. \label{good_sexp}
\end{equation}
This regularisation and the resulting flow  are plotted in Fig.~\eqref{fig:improved_blobs}d-f. { This flow is continuous and infinitely differentiable for $r>0$. The fluid stress is therefore also smooth and higher order flow singularities to be directly computed from it. However at $r=0$ the flow cannot be sampled directly because $(1-e^{-r/\epsilon})/r$ is undefined at this point. This is overcome by using the limit as $r\to 0$, which is well defined, instead of $r=0$.} Importantly when this flow is written in the sphere representation, Eq.~\eqref{eqn_Sphere_Model}, the dipole and the isotropic components are found to decay as $e^{-r/\epsilon}$. 

\subsection{An improved power-law regularisation}

The regularisations presented above converge rapidly in $\epsilon$ and can be used in order to obtain accurate computations. However, { power-law regularisation are popular \cite{Cortez,Cortez2015,Montenegro-Johnson2016,Montenegro-Johnson2015, Montenegro-Johnson2018,Godinez2015, Smith2018,Olson2013,Montenegro-Johnson2012,Smith}, continuously differentiable and can be directly sampled at $r=0$. Hence} it should be discussed how these blobs could be improved. Similarly to the exponential and the compact blobs the source dipole contribution can be removed by taking a linear combination of two power-law blobs. For example, the choice of regularisation
\begin{equation}\label{eq:52}
f_{\epsilon}^{p_{i}}(r) = \frac{105 \epsilon^{6}}{16 \pi r_\epsilon^{9}} - \frac{15 \epsilon^{4}}{ 8 \pi r_{\epsilon}^{7}}
\end{equation}
produces the flow tensor
\begin{equation}\label{eq:53}
\mathbf{S}_{\epsilon}^{p_{i}}(\mathbf{r}) = \frac{\mathbf{I} (2 r^{4} + 5 r^{2} \epsilon^{2} + 6 \epsilon^{4}) + \mathbf{r} \mathbf{r} ( 2 r^{2} + 5 \epsilon^{2}) }{16 \pi r_{\epsilon}^{5}}
,\end{equation}
where $r_\epsilon=\sqrt{r^2+\epsilon^2}$. A Taylor expansion of this flow shows that the first non-zero flow contribution, after the stokeslet, occurs at $\epsilon^{4}$. This contribution arises from the body forces generated by $1/r_{\epsilon}^{7}$, as predicted in the general expansion. The addition of higher powers of $r_{\epsilon}$ cannot remove this contribution since it is created by the first term in the Laurent expansion of $1/r_{\epsilon}^{7}$. The convergence of the flow from a power blob of the form $f_{\epsilon}(r) = A/r_{\epsilon}^{n+5}+B/r_{\epsilon}^{n+7}$ is therefore proportional to $\epsilon^{n+2}$, where $A$ and $B$ are constants chosen to normalise the blob and satisfy Eq.~\eqref{0dip}. Hence higher powers produce more accurate flows but are always polynomial in accuracy.

\section{Conclusion and future work} \label{sec:conclusion}
In this paper we analysed the flow from an arbitrary regularised stokeslet in order  to better understand the accuracy  numerical methods based on its use. The flow far from the centre of a general regularised stokeslet is shown to be composed of an infinite number of singularity and non-singularity flows weighted by powers of $\epsilon$, the ``size'' of the regularisation blob. The non-singularity solutions occur in blobs with power-law or slower decays and arise from a non-zero body force on the fluid. These leading non-singularity solution can arise at $\epsilon^{n+1}$ for regularisation blobs that decay like $f(r) \sim 1/r^{n+4}$. Conversely, the far-field flow from exponentially decaying blobs or faster are always composed of singularity solutions. This behaviour is reflected by the three example blobs considered: a popular power-law blob, a known Gaussian blob, and a blob with compact support.

As the characteristic size of the regularisation blob reduces to zero, all singularity and non-singularity terms above the stokeslet vanish. This is, of course,  a fundamental requirement for the regularised stokeslet technique. The rate at which the flow converge to the stokeslet, however, depends on the symmetries of the regularisation blob. Blobs without any symmetry will induce force dipole flows proportional to $\epsilon$ in the far-field and thus approach the real flow linearly with $\epsilon$ as $\epsilon \to 0$. On the other hand, if the blob has three symmetry planes all the additional singularities and non-singularity flows must be even powers of $\epsilon$ and so the real stokeslet flow is recovered at a rate $\epsilon^{2}$ as $\epsilon \to 0$. In the special case of spherically symmetric blobs, the infinite number of flow structures is found to truncate to a point force and source dipole for quickly decaying blobs. This source dipole contribution cannot be removed with a blob that is always greater than 0, and is proportional to $\epsilon^{2}$. Strictly positive spherically symmetric regularisation are by far the most common used regularisations in numerical simulations and therefore real flows are always recovered at the rate $\epsilon^{2}$ as $\epsilon \to 0$ in this case.

The complete flow from any spherically-symmetric blobs can further be written in terms of a point force, a source dipole and an isotropic component multiplied by functions of the distance from the origin. This representation relates the force distribution of the blob to the surface traction exerted by a translating sphere. This representation provides a simple method to compute the far-field flow structures and reveals that the flow near the centre of spherically-symmetric regularised stokeslets become isotropic, in contrast with the inherent anisotropy of Stokes flows and stokeslets. The regularisation parameter $\epsilon$ must therefore be kept much smaller than all other lengths to minimise the influence of this region in any numerical implementation. The contribution from the near-centre isotropic flow component reduces with $\epsilon$ for boundary integral representations but is a constant for some regularised slender-body theory representations \cite{Smith}. These slender-body theories use $\epsilon$ as a physical length and so this contribution should remain non-negligible. Furthermore, the leading flow around these slender structures is known to be that of a stokeslet and a source dipole \cite{1976} and so are best captured by blobs with compact support or exponential decays.

The results of our flow analysis can be used to develop regularisations that converge to the stokeslet flow exponentially, or at least {as fast}, with $\epsilon$ as $\epsilon \to 0$. { These regularisations are spherically symmetric and contain regions of negative force to remove the $\epsilon^{2}$ contribution to the flow.} We developed two new  regularisations with these properties: a compact support blob given by Eqs.~\eqref{good_fsup} and \eqref{good_ssup}, and an exponentially decaying blob given by Eqs.~\eqref{good_fexp} and \eqref{good_sexp}. Both these regularisation rapidly recover the flow form a stokeslet. This significantly reduces the error created by regularisation and so they are ideal for numerical studies. A method to increase the accuracy of power-law blobs beyond the $\epsilon^{2}$ limit was also discussed with the resultant flow always converging with a power of $\epsilon$ (Eqs.~\ref{eq:52} and \ref{eq:53}).

In addition to revealing  physical properties and limitations of the regularised stokeslet representation, our work creates avenues for future explorations and optimisations of regularised stokeslets. Similar investigations are needed to understand the flows from regularised point forces near boundaries \cite{Ainley2008,Cortez2015}, regularised higher-order singularities \cite{Cortez2015} or in Brinkman fluids \cite{Nguyen2018}. The identification of the flow could also help with the design of regularised stokeslets for specific purposes, such as optimal regularisations for densely packed systems, or regularisations to reflect inherent geometric asymmetries. From a more theoretical point of view, it would also be interesting to know if other flow-geometric parallel representations exist similarly to the relationship derived between spherically symmetric blobs and translating spheres.

\acknowledgements
This project has received funding from the European Research Council (ERC) under the European Union's Horizon 2020 research and innovation programme  (grant agreement 682754 to EL). BZ was supported by Trinity College, Cambridge for this project.

\appendix

\bibliographystyle{ieeetr}
\bibliography{UnderstandingRegularizedStokesletsBib}

\end{document}